\documentclass[aps,prd,twocolumn,groupedaddress,superscriptaddress,amsmath,amssymb, floatfix]{revtex4-2}
\usepackage{lineno}
\usepackage{graphicx} 
\usepackage{color}
\usepackage{hyperref}

\begin{document}
\bibliographystyle{apsrev4-2}
\bibliographystyle{h-physrev4}

\title{First measurement of the strange axial coupling constant using neutral-current quasielastic interactions of atmospheric neutrinos at KamLAND}

\newcommand{\tohoku}{\affiliation{Research Center for Neutrino Science, Tohoku University, Sendai 980-8578, Japan}}
\newcommand{\ipmu}{\affiliation{Kavli Institute for the Physics and Mathematics of the Universe (WPI), The University of Tokyo Institutes for Advanced Study, 
    The University of Tokyo, Kashiwa, Chiba 277-8583, Japan}}
\newcommand{\tohokuphys}{\affiliation{Department of Physics, Tohoku University, Sendai, 980-8578, Japan}}
\newcommand{\obihiro}{\affiliation{Department of Human Science, Obihiro University of Agriculture and Veterinary Medicine, Obihiro 080-8555, Japan}}
\newcommand{\butsuryo}{\affiliation{Faculty of Health Sciences, Butsuryo College of Osaka, Osaka, Japan}}
\newcommand{\osaka}{\affiliation{Graduate School of Science, Osaka University, Toyonaka, Osaka 560-0043, Japan}}
\newcommand{\rcnp}{\affiliation{Research Center for Nuclear Physics, Osaka University, Ibaraki, Osaka 567-0047, Japan}}
\newcommand{\tokushima}{\affiliation{Department of Physics, Tokushima University, Tokushima 770-8506, Japan}}
\newcommand{\tokushimags}{\affiliation{Graduate School of Integrated Arts and Sciences, Tokushima University, Tokushima 770-8502, Japan}}
\newcommand{\lbl}{\affiliation{Nuclear Science Division, Lawrence Berkeley National Laboratory, Berkeley, California 94720, USA}}
\newcommand{\hawaii}{\affiliation{Department of Physics and Astronomy, University of Hawaii at Manoa, Honolulu, Hawaii 96822, USA}}
\newcommand{\mitech}{\affiliation{Massachusetts Institute of Technology, Cambridge, Massachusetts 02139, USA}}
\newcommand{\ud}{\affiliation{University of Delaware, Newark, Delaware 19716, USA}}
\newcommand{\ut}{\affiliation{Department of Physics and Astronomy, University of Tennessee, Knoxville, Tennessee 37996, USA}}
\newcommand{\tunl}{\affiliation{Triangle Universities Nuclear Laboratory, Durham, 
    North Carolina 27708, USA; \\
    Physics Departments at Duke University, Durham, North Carolina 27708, USA; \\
    North Carolina Central University, Durham, North Carolina 27707, USA; \\
    and The University of North Carolina at Chapel Hill, Chapel Hill, North Carolina 27599, USA}}
\newcommand{\vt}{\affiliation{Center for Neutrino
   Physics, Virginia Polytechnic Institute and State University, Blacksburg,
   Virginia 24061, USA}}
\newcommand{\washington}{\affiliation{Center for Experimental Nuclear Physics and Astrophysics, 
    University of Washington, Seattle, Washington 98195, USA}}
\newcommand{\nikhef}{\affiliation{Nikhef and the University of Amsterdam, 
    Science Park, Amsterdam, the Netherlands}}
\newcommand{\tokyo}{\affiliation{Center for Nuclear Study, The University of Tokyo, 
    Tokyo 113-0033, Japan}}
\newcommand{\gppu}{\affiliation{Graduate Program on Physics for the Universe, Tohoku University, Sendai 980-8578, Japan}}
\newcommand{\frontier}{\affiliation{Frontier Research Institute for Interdisciplinary Sciences, Tohoku University, Sendai, 980-8578, Japan}}
\newcommand{\bu}{\affiliation{Boston University, Boston, Massachusetts 02215, USA}}
\newcommand{\chapel}{\affiliation{UNC Physics and Astronomy, 120 E. Cameron Ave., Phillips Hall CB3255, Chapel Hill, NC 27599}}

\newcommand{\aticrrnow}{\altaffiliation
    {Present address: Kamioka Observatory, Institute for Cosmic-Ray Research, 
    The University of Tokyo, Hida, Gifu 506-1205, Japan}}
\newcommand{\atqstnow}{\altaffiliation
    {Present address: National Institutes for Quantum and Radiological Science 
    and Technology (QST), Sendai 980-8579, Japan}}

%
%
\author{S.~Abe}\tohoku
\author{S.~Asami}\tohoku
\author{M.~Eizuka}\tohoku
\author{S.~Futagi}\tohoku
\author{A.~Gando}\tohoku
\author{Y.~Gando}\tohoku\obihiro
\author{T.~Gima}\tohoku
\author{A.~Goto}\tohoku
\author{T.~Hachiya}\tohoku
\author{K.~Hata}\tohoku
\author{K.~Ichimura}\tohoku
\author{S.~Ieki}\tohoku
\author{H.~Ikeda}\tohoku
\author{K.~Inoue}\tohoku
\author{K.~Ishidoshiro}\tohoku
\author{Y.~Kamei}\tohoku
\author{N.~Kawada}\tohoku
\author{Y.~Kishimoto}\tohoku\ipmu
\author{M.~Koga}\tohoku\ipmu
\author{M.~Kurasawa}\tohoku
\author{T.~Mitsui}\tohoku
\author{H.~Miyake}\tohoku
\author{T.~Nakahata}\tohoku
\author{K.~Nakamura}\tohoku
\author{R.~Nakamura}\tohoku
\author{H.~Ozaki}\tohoku\gppu
\author{T.~Sakai}\tohoku
\author{I.~Shimizu}\tohoku
\author{J.~Shirai}\tohoku
\author{K.~Shiraishi}\tohoku
\author{A.~Suzuki}\tohoku
\author{Y.~Suzuki}\tohoku
\author{A.~Takeuchi}\tohoku
\author{K.~Tamae}\tohoku
\author{H.~Watanabe}\tohoku
\author{Y.~Yoshida}\tohoku
\author{S.~Obara}\atqstnow\frontier
\author{A.K.~Ichikawa}\tohokuphys

\author{S.~Yoshida}\osaka

\author{S.~Umehara}\rcnp

\author{K.~Fushimi}\tokushima
\author{K.~Kotera}\tokushimags
\author{Y.~Urano}\tokushimags

\author{B.E.~Berger}\ipmu\lbl
\author{B.K.~Fujikawa}\ipmu\lbl

\author{J.G.~Learned}\hawaii
\author{J.~Maricic}\hawaii

\author{S.N.~Axani}\ud

\author{Z.~Fu}\mitech
\author{J.~Smolsky} \mitech
\author{L.A.~Winslow}\mitech

\author{Y.~Efremenko}\ipmu\ut

\author{H.J.~Karwowski}\tunl
\author{D.M.~Markoff}\tunl
\author{W.~Tornow}\ipmu\tunl

\author{J.A.~Detwiler}\ipmu\washington
\author{S.~Enomoto}\ipmu\washington

\author{M.P.~Decowski}\ipmu\nikhef

\author{C.~Grant}\bu
\author{A.~Li}\bu\tunl
\author{H.~Song}\bu

\author{S.~Dell'Oro}\vt
\author{T.~O'Donnell}\vt

\collaboration{KamLAND Collaboration}\noaffiliation

\date{\today}

\begin{abstract}
We report a measurement of the strange axial coupling constant $g_A^s$ using atmospheric neutrino data at KamLAND.
This constant is a component of the axial form factor of the neutral-current quasielastic (NCQE) interaction.
The value of $g_A^s$ significantly changes the ratio of proton and neutron NCQE cross sections.
KamLAND is suitable for measuring NCQE interactions as it can detect nucleon recoils with low-energy thresholds and measure neutron multiplicity with high efficiency.
KamLAND data, including the information on neutron multiplicity associated with the NCQE interactions,
makes it possible to measure $g_A^s$ with a suppressed dependence on the axial mass $M_A$, which has not yet been determined.
For a comprehensive prediction of the neutron emission associated with neutrino interactions,
we establish a simulation of particle emission via nuclear deexcitation of $^{12}$C, a process not considered in existing neutrino Monte Carlo event generators.
Energy spectrum fitting for each neutron multiplicity gives $g_A^s =-0.14^{+0.25}_{-0.26}$, which is the most stringent limit obtained using NCQE interactions without $M_A$ constraints.
The two-body current contribution considered in this analysis relies on a theoretically effective model and electron scattering experiments and requires future verification by direct measurements and future model improvement.
\end{abstract}

\maketitle

\section{Introduction} \label{sec:Introduction}
Various experiments have measured neutrino-nucleon interactions, and our understanding of these interactions gradually deepens.
Among many neutrino interaction channels, the  neutral-current quasielastic (NCQE) interaction contains fundamental and interesting information about nucleons.
The NCQE interaction, $\nu_l + N \rightarrow \nu_l +N$, where $N$ denotes either a proton or neutron, does not change the lepton charge between the initial and final states.
In contrast, the charged-current quasielastic (CCQE) interaction, $\nu_\mu + n \rightarrow \mu^- + p$, does.
The CCQE interaction only involves isovector weak currents, while the NCQE interaction is sensitive to isoscalar weak currents.
Therefore, searching for strange quarks existing as sea quarks in nucleons through their isoscalar contribution to the NCQE interaction is possible.
In experiments, one measures the strange axial coupling constant $g_A^s$, which is the strange axial form factor at four-momentum transfer squared $Q^2=0$.
Since the $Q^2$ dependence of the axial form factor is parametrized by an axial mass $M_A$, the measured value of $g_A^s$ generally depends on the value of $M_A$.
\par
The BNL E734 experiment performed the first measurement of $g_A^s$ using the NCQE interaction~\cite{PhysRevD.35.785,PhysRevC.48.761}.
They used accelerator neutrinos and measured the $\nu + p \rightarrow \nu + p$ and $\bar\nu + p \rightarrow \bar\nu+p$ differential cross sections as a function of $Q^2$.
They confirmed a strong positive correlation between $g_A^s$ and $M_A$.
They obtained $g_A^s = -0.15 \pm 0.07$ with the strong constraint of $M_A=1.061 \pm 0.026$\,GeV, the world average at the time.
In the 1970s and 1980s, various measurements from deuteron-target bubble chambers appeared to be consistent with obtained results of $M_A\sim1.0$\,GeV~\cite{Bodek2008}.
However, recent experiments using carbon and oxygen targets have found results as large as $M_A = 1.1 - 1.3$\,GeV, and the discrepancy has become an issue~\cite{Kakorin2021}.
It is becoming clear that a two-body current contribution, called two-particle two-hole (2p2h), must be considered to explain this discrepancy~\cite{PhysRevC.83.045501,PhysRevC.80.065501,Bodek2011}.
A direct measurement of the 2p2h interaction has not yet been realized, so there is a model-dependent uncertainty.
Therefore, it is difficult to determine a reasonable constraint on $M_A$.
\par
The MiniBooNE Collaboration measured the flux-averaged NCQE differential cross section~\cite{PhysRevD.82.092005}.
Assuming $M_A=1.35$\,GeV, obtained from their CCQE analysis~\cite{PhysRevD.81.092005}, they found $g_A^s = 0.08 \pm 0.26$.
In this analysis, they did not simultaneously fit $M_A$ and $g_A^s$.
Using the results provided by the MiniBooNE Collaboration, Golan {\it et al.} performed an independent simultaneous-fit analysis using the NuWro Monte Carlo event generator~\cite{PhysRevC.86.015505, PhysRevC.88.024612}.
This analysis also took into account the 2p2h contribution and obtained $M_A=1.10^{+0.13}_{-0.15}$\,GeV and $g_A^s=-0.4^{+0.5}_{-0.3}$, confirming a positive correlation between these parameters.
\par
The strange axial coupling constant $g_A^s$ corresponds to the strange quark-antiquark contribution to the nucleon spin, commonly represented by $\Delta s$.
Several experimental results have been obtained using polarized-lepton deep-inelastic scattering:
$\Delta s=-0.18 \pm 0.05$ from EMC~\cite{ASHMAN19891,ALBERICO2002227}, $\Delta s=-0.085 \pm 0.018$ from HERMES~\cite{PhysRevD.75.012007}, and $\Delta s=-0.08 \pm 0.02$ from COMPASS~\cite{ALEXAKHIN20078}.
These results rely on $SU(3)_f$ flavor symmetry.
The $SU(3)_f$ flavor symmetry is violated by a maximum of 20\%, in which case these results are shifted by $\pm 0.04$~\cite{ALEXAKHIN20078}.
This uncertainty is approximately equal to or larger than the statistical and systematic errors of the experiments mentioned above.
It is clearly of interest to measure $g_A^s (\Delta s)$ in a way that is independent of $SU(3)_f$ flavor symmetry, namely by measuring the NCQE interaction.
\par
One challenge in measuring $g_A^s$ using the NCQE interaction is the strong correlation with $M_A$.
In the BNL E734 and MiniBooNE experiments, a proton target was used primarily because of the difficulty of measuring NCQE on a neutron target.
The value of $g_A^s$ significantly changes the ratio of proton and neutron NCQE cross sections.
Therefore, a measurement exclusively on a proton target (or neutron target) depends highly on $M_A$ and other normalization uncertainties.
Conversely, when measuring the ratio, the normalization cancels out, and we can measure $g_A^s$ with only a slight dependence on $M_A$.
In practice, nucleons measured by detectors are affected by final-state interactions (FSI), nuclear deexcitation, and secondary interactions (SI).
These effects somewhat smear the information about the target nucleon.
Nevertheless, information about the target nucleons and $g_A^s$ can be extracted by measuring the neutron multiplicity with high efficiency.
\par
This paper aims to measure the neutron multiplicity of atmospheric neutrino NCQE interactions at KamLAND and to obtain $g_A^s$ with a slight dependence of $M_A$.
In addition to the 2p2h contribution, the nuclear deexcitation process, which can emit neutrons, is considered in our analysis.
Our paper is organized as follows:
Sec.~\ref{sec:Formalism} describes the formalism of the NCQE interaction;
Sec.~\ref{sec:deex} introduces a simulation of particle emission via nuclear deexcitation;
Sec.~\ref{sec:KamLAND} describes the KamLAND detector and data analysis;
Sec.~\ref{sec:Simulation} contains details of the Monte Carlo simulation;
Sec.~\ref{sec:Analysis_Results}, the analysis method and results;
and our conclusions are presented in Sec.~\ref{sec:Conclusion}.

\section{Formalism of neutral-current quasielastic Interaction} \label{sec:Formalism}
The Llewellyn-Smith formula~\cite{LLEWELLYNSMITH1972261} is commonly used to describe CC and NC QE interactions.
The hadronic current is composed of vector and axial parts.
Assuming a dipole form, the axial form factors of CC and NC are expressed as follows:
\begin{align} \label{eq:axial_form_factor}
  G_A^{\text{CC}}(Q^2)     &= g_A \left( 1+\frac{Q^2}{M_A^2} \right)^{-2}, \\
  G_A^{\text{NC},p/n}(Q^2) &= \frac{1}{2} (\pm g_A -g_A^s) \left( 1+\frac{Q^2}{M_A^2} \right)^{-2},
  \label{eq:strange_axial_form_factor}
\end{align}
where $g_A$ denotes the axial coupling constant, and the sign $+(-)$ is for proton (neutron).
A value of $g_A=1.2723\pm0.0023$ is determined by nucleon $\beta$ decay experiments~\cite{PhysRevD.98.030001}.
The strange quark contribution $g_A^s$ only appears in the form factors of NC.
\par
The relationship between the vector and electromagnetic form factors can be written as follows:
\begin{align}
  F_{1,2}^{\text{CC}}(Q^2) & =  F_{1,2}^{p}(Q^2) - F_{1,2}^{n}(Q^2) \label{eq:VFF1}, \\
  F_{1,2}^{\text{NC},p(n)} (Q^2) = &
    \pm \frac{1}{2} F_{1,2}^{\text{CC}} (Q^2) - 2\sin^2\theta_W F_{1,2}^{p(n)}(Q^2)  \nonumber \\
     & - \frac{1}{2}F_{1,2}^{s}(Q^2), \label{eq:VFF2}
\end{align}
where $\theta_W$ is the Weinberg angle, and the indices $p$ and $n$ represent the proton and neutron, respectively.
The vector form factors for the proton and neutron ($F_{1,2}^{p(n)}$) can be written in terms of the electric $G_E$ and magnetic $G_M$ form factors:
\begin{align}
  F_1^{p(n)}(Q^2) & = \left( 1 + \frac{Q^2}{4M^2}\right)^{-1} \nonumber \\
        & \times\left[ G_E^{p(n)}(Q^2) + \frac{Q^2}{4M^2} G_M^{p(n)}(Q^2) \right], \\
  F_2^{p(n)}(Q^2) & = \left( 1 + \frac{Q^2}{4M^2}\right)^{-1} \left[ G_M^{p/n}(Q^2) - G_E^{p(n)}(Q^2) \right],
\end{align}
where $M$ is the average of the proton and neutron masses.
The electric and magnetic form factors, $G_E$ and $G_M$, are formulated from electron scattering data.
The dipole form was commonly adopted in the past, as in the axial form factors.
However, as the deviation from the dipole form became apparent, a more sophisticated parametrization, BBBA05~\cite{BRADFORD2006127}, has recently been used.
The strange vector form factor $F_{1,2}^{s}(Q^2)$ in Eq.~(\ref{eq:VFF2}) can be expressed assuming a dipole form:
\begin{align}
  F_{1}^s(Q^2) & = F_{1}^s Q^2  \left(1+\frac{Q^2}{4M^2}\right)^{-1} \left(1+\frac{Q^2}{M_V^2}\right)^{-2}, \\
  F_{2}^s(Q^2) & = F_{2}^s(0)  \left(1+\frac{Q^2}{4M^2}\right)^{-1} \left(1+\frac{Q^2}{M_V^2}\right)^{-2},
\end{align}
where the vector mass $M_V=0.84$\,GeV is determined by electron scattering experiments.
A global analysis of the polarized electron elastic-scattering experiments shows that the values of strange vector form factors are consistent with zero~\cite{PhysRevC.78.015207}.
Thus, we set $F_{1}^s=F_{2}^s(0) = 0$ in this analysis.
\par
Generally, the vector form factors can be precisely determined from high-statistics electron-scattering data.
In contrast, the axial form factors are uncertain because they can be measured only through neutrino interactions.
As can be seen from Eq.~(\ref{eq:strange_axial_form_factor}), the extraction of $g_A^s$, the purpose of this paper, depends on both $g_A$ and $M_A$.
Since $g_A$ is precisely determined, the uncertainty in $M_A$ is the larger issue.
\par
\begin{figure}[htb]
\includegraphics[width=1.0\columnwidth]{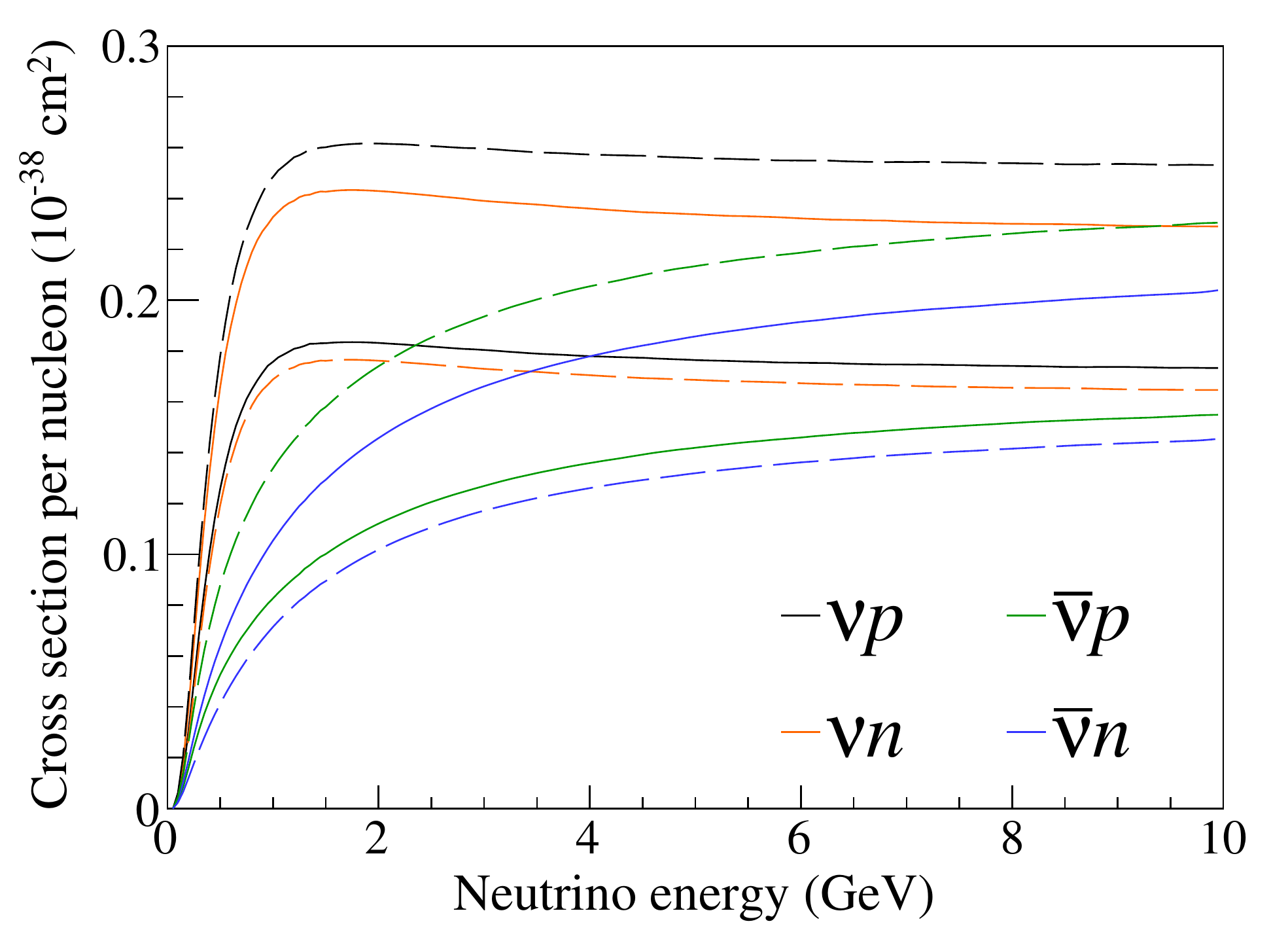}
\caption{NCQE cross section on carbon per nucleon as a function of neutrino energy.
         The black and orange lines represent the neutrino cross sections on protons and neutrons, respectively.
         The green and blue lines represent the antineutrino cross sections on protons and neutrons, respectively.
         The solid (dashed) lines are the cross sections with $g_A^s=0~(-0.3)$.
         These results are obtained using NuWro with $M_A=1.2$\,GeV~\cite{PhysRevC.86.015505}.
        }
\label{fig:xsec_NCQE}
\end{figure}
\begin{figure}[htb]
\includegraphics[width=1.0\columnwidth]{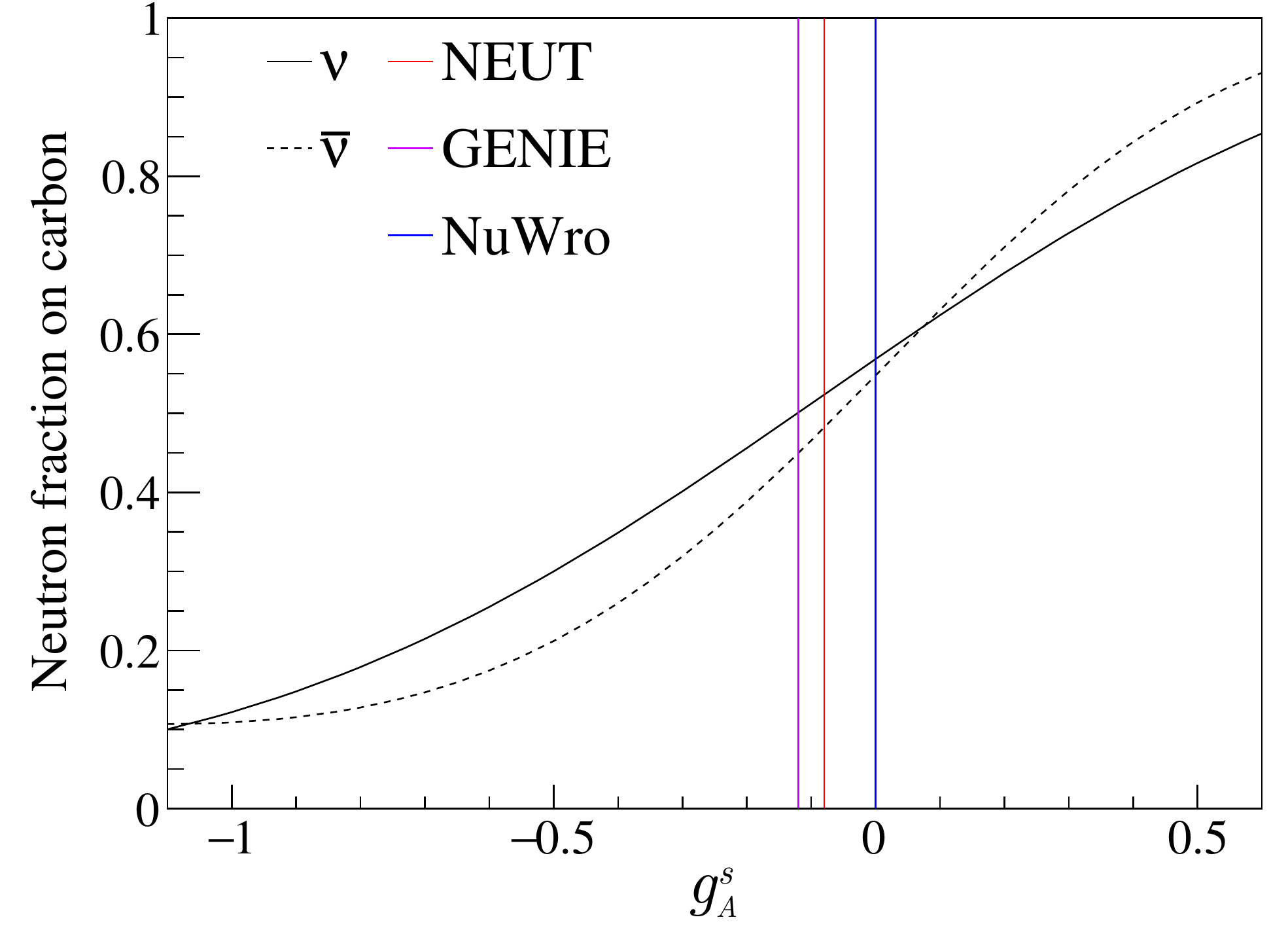}
\caption{Neutron fraction of the total NCQE cross section on carbon as a function of the strange axial coupling constant $g_A^s$.
         The solid (dashed) line represents the neutrino (antineutrino) cross-section fraction.
         This result is obtained using NuWro with $M_A=1.2$\,GeV at 0.5\,GeV neutrino energy.
         The red, violet, and blue vertical lines represent the default values adopted in neutrino Monte Carlo generators, NEUT~\cite{Hayato2021,PhysRevLett.108.052505}, GENIE~\cite{ANDREOPOULOS201087}, and NuWro~\cite{PhysRevC.86.015505}, respectively.
         Lower values of $g_A^s$ lead to a lower neutron contribution to the total cross section.
        }
\label{fig:xsec_NCQE_ratio}
\end{figure}

The strange axial coupling constant $g_A^s$ significantly changes the relative proton and neutron NCQE cross sections with little change in the total cross section.
Figure~\ref{fig:xsec_NCQE} shows the NCQE cross section on carbon per nucleon in NuWro~\cite{PhysRevC.86.015505}.
For lower values of $g_A^s$, the neutron contribution to the total cross section becomes smaller while the proton contribution increases.
This trend is also evident by looking at the neutron cross section as a fraction of the total NCQE cross section, as shown in Fig.~\ref{fig:xsec_NCQE_ratio}.
The value of $M_A$ changes the shape of the NCQE differential cross section and the overall cross section normalization.
These changes are almost equal for proton- and neutron-target cross section contributions.
Therefore, measurements of only the proton-target (or neutron-target) NCQE interaction depend highly on uncertainties in normalization factors such as $M_A$ and the neutrino flux.
In contrast, measuring the neutron-target cross section as a fraction of the total NCQE cross section makes it possible to measure $g_A^s$ with less dependence on these normalization factors.
The nucleons measured by detectors are affected by FSI and SI, so it is impossible to strictly identify the target nucleons on an event-by-event basis.
However, by measuring nucleon multiplicity, it is possible to statistically separate the contribution of target nucleons using the distribution, within the uncertainty of these nuclear effects.
This method requires high nucleon detection efficiency.
In this analysis, we measured neutron multiplicity using KamLAND, which has a neutron detection efficiency of over 80\%.
\par
The default values of $g_A^s$ adopted in common neutrino-interaction Monte Carlo generators are different:
$g_A^s=-0.08$ in NEUT version 5.4.0.1~\cite{Hayato2021,PhysRevLett.108.052505},
$g_A^s=-0.12$ in GENIE version 3.00.06~\cite{ANDREOPOULOS201087},
and $g_A^s=0$ in NuWro version 21.09.
These differences change the neutron fraction of the total NCQE cross section on carbon by about 10\%.

\section{Nuclear deexcitation associated with neutrino-nucleus interaction} \label{sec:deex}
Nuclear deexcitation often occurs associated with neutrino-nucleus interactions.
The typical excitation energy is about 20\,MeV in the case of the $^{12}$C target~\cite{PhysRevD.67.076007}.
The excitation energy is higher than the separation energies of various particles, including neutrons, protons, and $\alpha$ particles.
Various types of particles can therefore be emitted via deexcitation processes.
It is important to predict these nuclear processes, especially for experiments measuring neutron multiplicity.
However, current sophisticated neutrino Monte Carlo event generators, such as NuWro, NEUT, and GENIE, do not take them into account.
Here, we have established a systematic method to predict nuclear deexcitation~\cite{Abe_2021}.
This method can be used with the results of neutrino Monte Carlo event generators.
Since this study is intended for use in liquid scintillator detectors, including KamLAND, we only discuss the $^{12}$C target.

\subsection{Overview of the prediction}
Neutrino Monte Carlo event generators are event-by-event simulations, so we need an event-by-event deexcitation model to use them.
We use two simulation software packages in this prediction, TALYS version 1.95~\cite{KONING20122841} and a modification of Geant4 version 10.7.p03~\cite{AGOSTINELLI2003250}.
\par
TALYS is an open-source software package for the simulation of nuclear reactions.
It provides a complete and accurate nuclear reaction simulation up to 200\,MeV, including fission, scattering, and compound reactions.
Given any nucleus and excitation energy, it provides the branching ratios of all nuclear deexcitation processes.
Although TALYS provides branching ratios, it does not perform event-by-event simulations.
\par
Geant4, a widely-used software package for simulating the passage of particles through matter, makes it possible to do the event-by-event simulation.
Within Geant4, ``G4RadioactiveDecay'' simulates nuclear deexcitation and radioactive decay.
An event-by-event simulation of deexcitation decay chains is performed by loading the branching ratios obtained from TALYS into G4RadioactiveDecay with several modifications.
In addition to the branching ratios from TALYS, various parametrizations related to the shell model, including excitation energies and spectroscopic factors, are necessary for the simulation.

\subsection{Shell model picture of $^{12}$C}\label{sec:shell_model}
In the simple shell model picture of the $^{12}$C ground state, two nucleons lie in the $s_{1/2}$ shell, four nucleons lie in the $p_{3/2}$ shell, and no nucleon lies in the $p_{1/2}$ shell.
When a nucleon in the $p_{3/2}$ shell disappears, the excitation energy is zero, leading to no deexcitation.
Assuming the same probability for all nucleons, the spectroscopic factors for $s_{1/2}$ and $p_{3/2}$ are 1/3 and 2/3, respectively.
However, it is known that the actual spectroscopic factor of $s_{1/2}$ is smaller than this value because it is more tightly bound than $p_{3/2}$.
We adopt 0.296 for $s_{1/2}$ and 0.704 for $p_{3/2}$ from electron scattering data~\cite{PhysRevC.61.064325}.

\subsection{Disappearance from the $p$ shell}\label{sec:dis_p}
In a more precise shell model picture, the $p_{1/2}$ shell is partially occupied by a nucleon pair due to nucleon-nucleon correlation.
From various shell model calculations, this partial occupation, called the pairing effect, is expected to occur with a probability of $40\pm10$\%~\cite{PhysRevD.67.076007}.
Therefore, $20\pm5$\% of the time, the disappearance of a single nucleon from the $p_{3/2}$ or $p_{1/2}$ shell will leave the residual nucleus, $^{11}$C or $^{11}$B, in an excited state with spin-parity $J^{\pi}=1/2^-$. 
The energy gap between $p_{1/2}$ and $p_{3/2}$ is a few MeV.
There is only one excited state in both $^{11}$C and $^{11}$B.
It decays to the ground state by emitting one $\gamma$ with an energy of 2.0\,MeV for $^{11}$C and 2.1\,MeV for $^{11}$B.

\subsection{Disappearance from the $s_{1/2}$ shell}\label{sec:dis_s}
Nucleon disappearance from the $s_{1/2}$ shell is more complicated than disappearance from the $p$ shell.
Because of the high excitation energy, typically more than the separation energies, we need to consider various particle emissions, including multistep processes as well as single-step deexcitations.
The branching ratios for $\gamma, \alpha, n, p$, deuteron ($d$), triton ($t$), and $^3$He emissions are extracted from TALYS, including the full decay chains of the daughter nuclei.
Since the excitation energy of an $s_{1/2}$-hole is large, the impact of the pairing effect is negligible.
\par
\begin{figure}[htb]
\centering
\includegraphics[width=1.0\columnwidth]{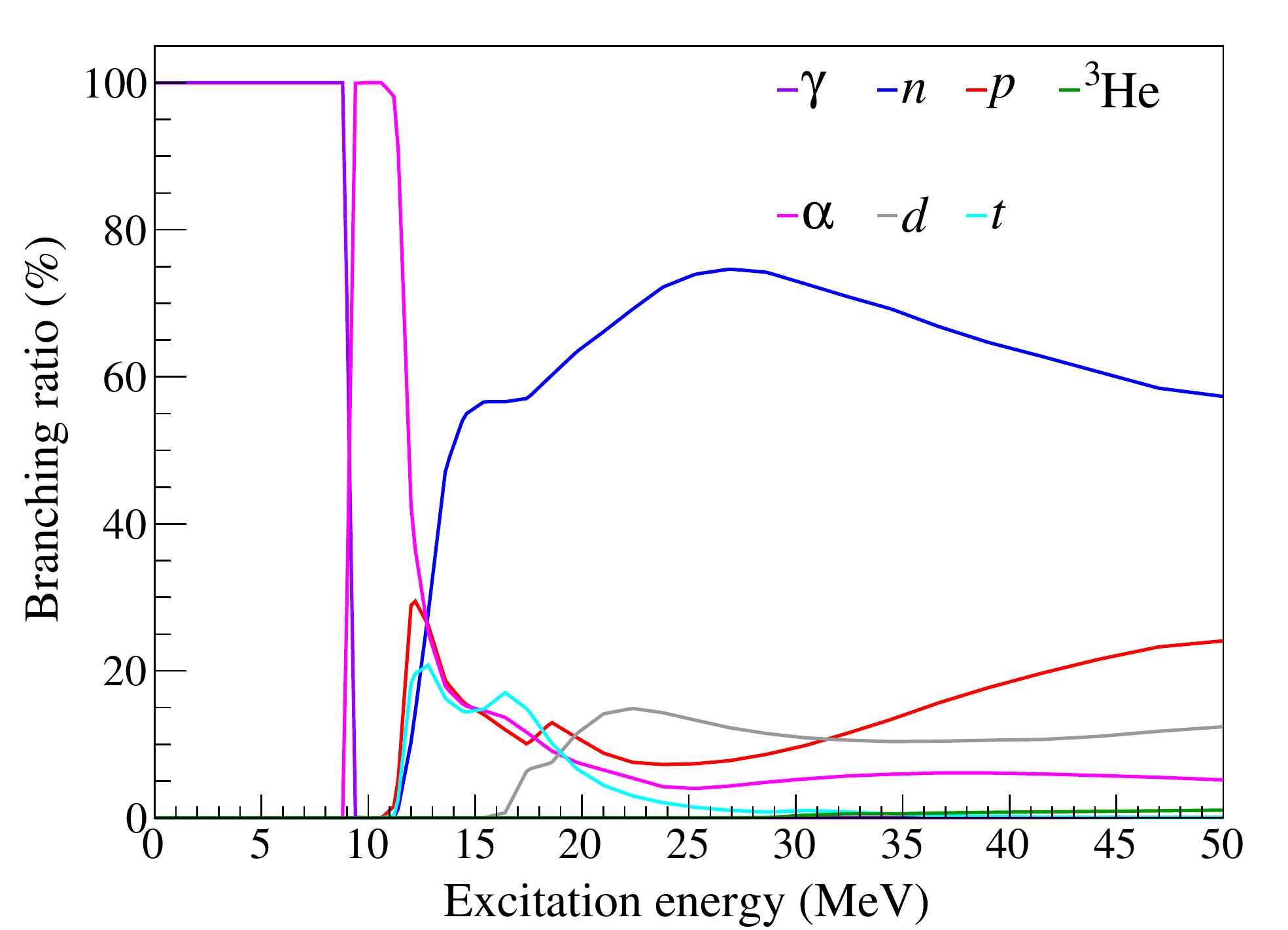}
\caption{Branching ratios of $^{11}$B$^*$ $J^\pi=1/2^+$ deexcitations as a function of excitation energy calculated with TALYS~\cite{KONING20122841}.
         At a typical excitation energy of 23\,MeV, neutron emission accounts for about $65$\% of decays. 
        }
\label{fig:br_11B}
\end{figure}

Fig.~\ref{fig:br_11B} shows the branching ratios of $^{11}$B$^*$ decay as a function of excitation energy calculated with TALYS.
The spin-parity is $J^\pi=1/2^+$ for single nucleon disappearance from the $s_{1/2}$ shell.
At the typical excitation energy of 23\,MeV, neutron emission accounts for about $65$\% of deexcitations.
This process strongly affects the neutron multiplicity associated with neutrino-nucleus interactions.
In contrast, the neutron branching ratio for $^{11}$C$^*$ decay at a 23\,MeV excitation energy is about 6\%.
This branching ratio is similar to that of proton emission for $^{11}$B$^*$.

The excitation energy of $s_{1/2}$-hole state has a finite width and is commonly parametrized with a Lorentzian distribution.
We adopt $E=23\pm1$\,MeV as the mean and $\Gamma=14^{+10}_{-2}$\,MeV as the FWHM width from electron scattering data~\cite{PhysRevC.61.064325,NAKAMURA1976381}.
We briefly mention how the uncertainty of these values affects the branching ratios at the end of this section.
\par
We simulate the deexcitation decay chain event by event with Geant4 using branching ratios extracted from TALYS and the excitation energy distribution.
The original Geant4 code does not treat emissions of tritons, deuterons, or $^3$He, so we modified the code to implement these decay modes.
The kinematics of the deexcitation process, such as separation energies and recoil, is taken into account properly.

\subsection{Comparison with experimental data and other predictions}
\begin{figure}[htb]
\centering
\includegraphics[width=1.0\columnwidth]{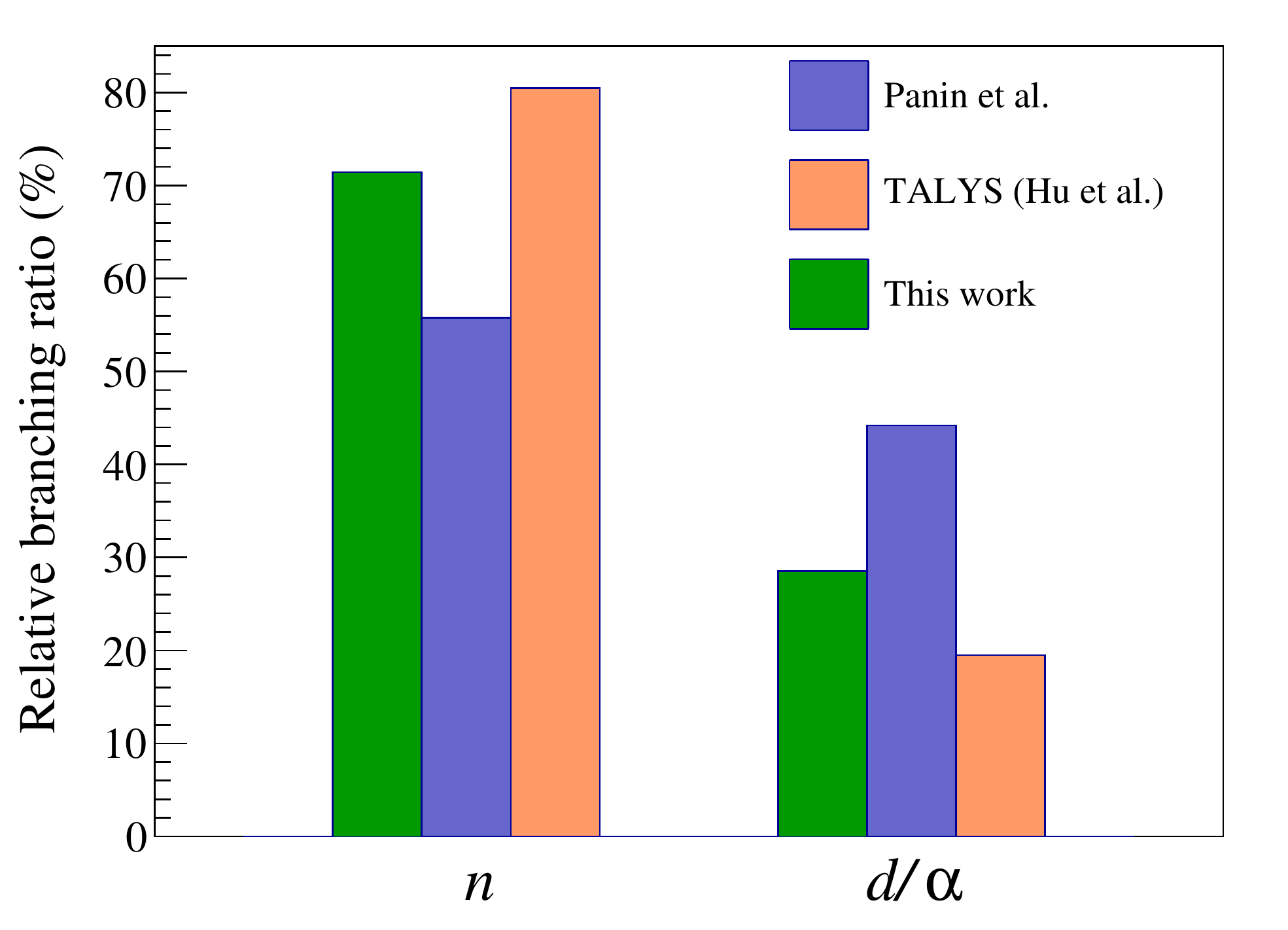}
\caption{Comparison of measured and predicted relative branching ratios of $n$ and $d/\alpha$ for $^{11}$B$^*$ with 16-35\,MeV excitation energy.
         The experimental data, in blue, are from Panin {\it et al.}~\cite{PANIN2016204}.
         The orange histograms show the predicted results from Hu {\it et al.} using TALYS~\cite{HU2022137183},
         and the greens represent our results.
         }
\label{fig:br_nda}
\end{figure}

We compare our prediction with experimental data and other predictions.
Figure~\ref{fig:br_nda} shows a comparison of the relative branching ratios of $n$ and $d/\alpha$ for $^{11}$B$^*$ with excitation energies of 16-35\,MeV.
The experimental data are from Panin {\it et al.}~\cite{PANIN2016204}, which measured three single-step deexcitation modes:
$^{11}$B$^* \rightarrow n + ^{10}$B,
$^{11}$B$^* \rightarrow d + ^{9}$Be, and
$^{11}$B$^* \rightarrow \alpha + ^{7}$Li.
The published result does not distinguish between $d$ and $\alpha$,
so for comparison, we calculate the relative branching ratios of $n$ and $d/\alpha$.
Another prediction result from Hu {\it et al.} uses TALYS version 1.95~\cite{HU2022137183}, the same version used in our analysis.
The excitation energy and spin-parity configurations may cause the difference between Hu's result and ours.
The branching ratio to $n$ is the most important parameter in this analysis.
Our result agrees with the experimental data within a relative uncertainty of 20\%.

\begin{figure}[htb]
\centering
\includegraphics[width=1.0\columnwidth]{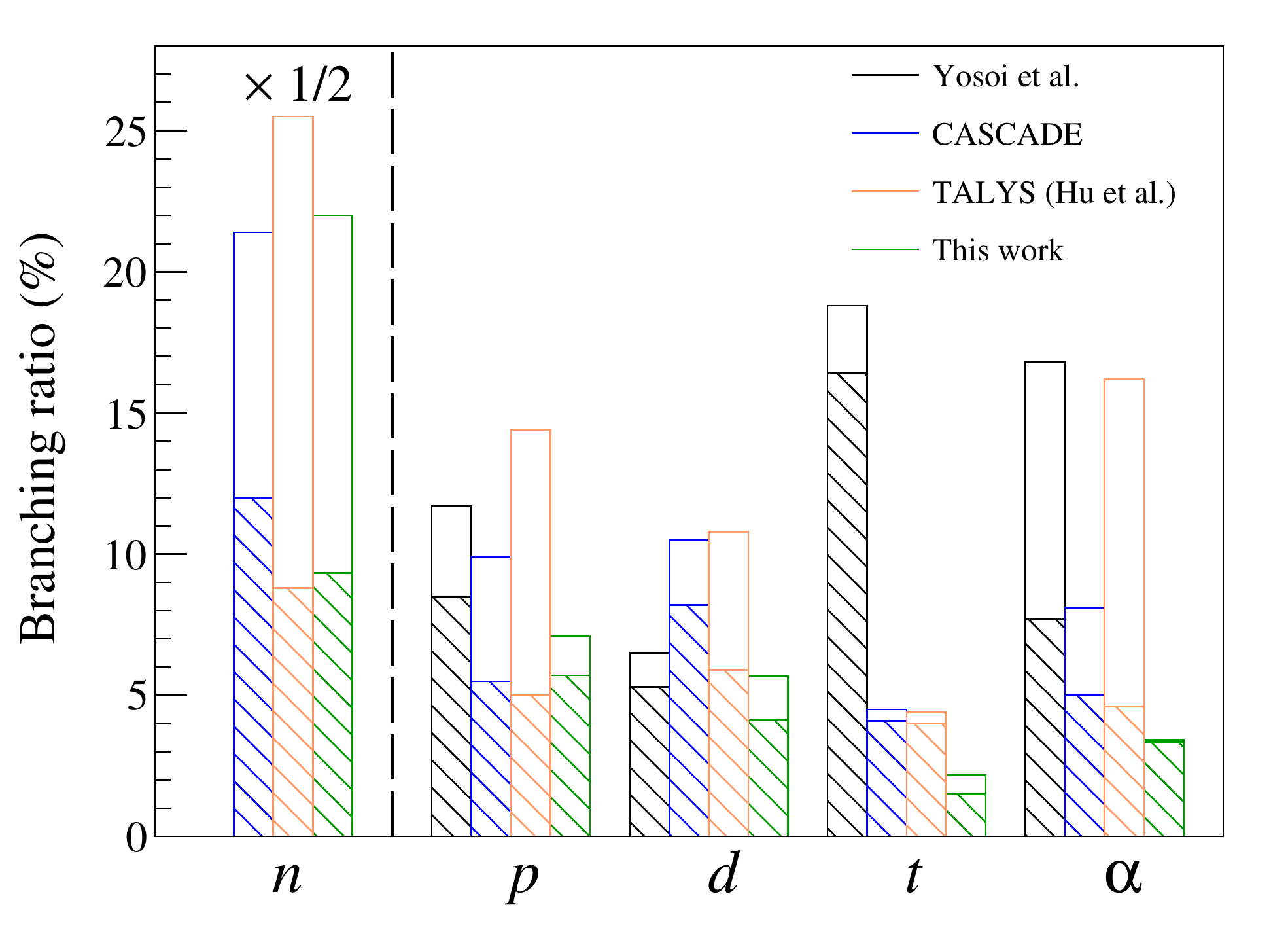}
\caption{Comparison of measured and predicted branching ratios of $n, p, d, t,$ and $\alpha$ for $^{11}$B$^*$ with 16-35\,MeV excitation energy.
         The branching ratios of $n$ are multiplied by a factor of 1/2.
         The green histograms represent our result using TALYS, and the orange histograms represent the prediction by Hu {\it et al.} using TALYS~\cite{HU2022137183}.
         The experimental data in black are from Yosoi {\it et al.}, and the authors also provide the predicted result using the CASCADE code~\cite{YOSOI2003255}.
         The hatched histograms represent the branching ratios for single-step decays, and the open histograms represent those from multistep decays.
        }
\label{fig:br_compare}
\end{figure}

Figure~\ref{fig:br_compare} compares the measured and predicted branching ratios for $^{11}$B$^*$ in the same excitation energy range with the experimental result from Yosoi {\it et al.}~\cite{YOSOI2003255}.
The $^3$He branching ratio is not shown because it is less than 1\%.
The $n$ branching ratios are consistent within a 20\% relative uncertainty.
There is a large difference in the single-step decay of triton, where the experimental result has a much larger value than the predictions.
It is seen from Fig.~\ref{fig:br_11B} that such a high branching ratio can not be explained by the model implemented in TALYS.
The authors also discussed this issue, but the causes are still unclear.
Further checks are needed, such as validation experiments and model evaluations.
We also confirmed a large difference in the multistep $\alpha$ decay.
Our result gives almost 0\% while others show about 5\%.
The $\alpha$ emission process is dominant at low excitation energies around 10\,MeV, which lead to low $\alpha$ kinetic energies and low excitation energies of the daughter nuclei.
Since the neutron separation energy of $^7$Li is as high as 7.3\,MeV, multistep $\alpha$ deexcitations do not contribute significantly to neutron emission.
All these differences between our prediction and experimental results and with other predictions are considered model-dependent uncertainties.

We also compare the branching ratios of $^{11}$C$^*$ with another prediction by Kamyshkov {\it et al.} using SMOKER~\cite{PhysRevD.67.076007}.
The SMOKER code does not consider the deexcitation modes of $d, t,$ and $^3$He, which account for about 15\% of the total.
We therefore only compare the $n, p,$ and $\alpha$ branching ratios.
In contrast to $^{11}$B$^*$, neutron emission is a minor deexcitation mode, while proton emission is a major one.
The total branching ratio for single-step and multistep neutron decays is 5.7\%, while SMOKER predicts 13.8\%.
This difference is also treated as a model-dependent uncertainty.
\par
Finally, we check the impact of the mean and width of the excitation energy distribution on the branching ratios.
The relative changes in the branching ratios are within 15\% when each parameter is changed within its uncertainty.
We assign this uncertainty from the excitation energy in addition to the model-dependent uncertainty derived from Figs.~\ref{fig:br_nda} and \ref{fig:br_compare}.

\section{KamLAND Data} \label{sec:KamLAND}
\subsection{KamLAND detector and data set}
KamLAND, a 1000-ton liquid-scintillator (LS) detector, is located 1000\,m underground in the Kamioka mine, Japan.
The cosmic muon flux is suppressed by a factor of $10^{-5}$ relative to sea level.
The detector consists of an 18\,m diameter stainless-steel spherical tank that defines the boundary of inner and outer detectors (ID and OD, respectively).
The inner surface of the tank is instrumented with 1325 17-inch photomultiplier tubes (PMTs) and 554 20-inch PMTs facing the center of the detector.
A 13\,m diameter EVOH/nylon balloon is suspended containing 1000\,tons of LS.
The elemental composition of the LS is approximately CH$_2$~\cite{PhysRevC.81.025807}.
The space between the balloon and the tank is filled with nonscintillating mineral oil, operating as a buffer (BO).
The OD is a cylindrical vessel filled with pure water.
This region is instrumented with 140 20-inch PMTs, acting as a cosmic-ray muon veto.
Further details of the detector are in~\cite{Suzuki2014}.
\par
The KamLAND data used in this paper are based on a total live time of 10.74\,years, acquired between January 2003 and May 2018.
The data set is divided into four periods: Periods I, II, III, and IV.
The major detector changes are as follows.
Period I (3.77\,years of live time) ended in May 2007, when we embarked on a LS purification campaign.
This purification work changed the scintillation and optical parameters, so we analyze this period separately.
Period II (1.79\,years of live time) started at the end of the purification campaign in April 2009 and ended in August 2011.
At that time, we started KamLAND-Zen 400 experiment by installing a 154-cm-radius inner balloon (IB) at the center of the KamLAND~\cite{PhysRevC.85.045504}.
Period III (3.66\,years of live time) refers to data during KamLAND-Zen 400 experiment from October 2011 to August 2015.
After this period, we extracted the IB and refurbished the OD system in 2016~\cite{Ozaki:2017TS}.
Period IV (1.52\,years of live time) started after the OD refurbishment in April 2016.
\par
After the end of period IV, we started KamLAND-Zen 800 by installing a new 190\,cm radius inner balloon.
The data acquired during KamLAND-Zen 800 is not included in this analysis.

\subsection{Event selection} \label{sec:selection}
KamLAND detects neutrino interactions via scintillation light.
There is no threshold for scintillation light, unlike Cherenkov light.
As a result, a scintillator detector like KamLAND can detect not only charged leptons and pions but also protons and neutrons with low-energy thresholds.
Protons directly produce scintillation light through ionization, while neutrons are detectable via proton recoils and later through capture on nuclei.
The primary energy deposition by the proton recoils occurs very quickly on the order of ns, while the capture has a much longer lifetime of several hundred $\mu$s, making it possible to perform delayed coincidence measurements.
Since the NCQE interaction mainly emits protons and neutrons, this feature of scintillator detectors makes it possible for us to measure the NCQE interaction.
\par
A neutrino interaction in KamLAND produces a prompt event caused by the energy deposit of charged particles and neutron recoils.
Neutrons are then captured by protons (or $^{12}$C) with a lifetime of $207.5 \pm 2.8\,\mu$s~\cite{PhysRevC.81.025807}, emitting a 2.2\,MeV (4.9\,MeV) gamma ray which produces a delayed event.
We can observe the neutron capture events with high accuracy by performing delayed coincidence measurements using time and spatial correlations of prompt and delayed events.
\par
We give some notes on the energy and vertex used in this paper.
We use visible energy to evaluate the atmospheric neutrino events here.
For CC events, the visible energy includes the energy deposit of the final-state lepton (electron or muon).
On the other hand, in the case of NC events, the visible energy does not include that of the final-state lepton (neutrino), leading to lower-prompt visible energy than CC events.
The vertex used in this paper is almost equivalent to the centroid of the energy deposition.
Since we cannot distinguish the energy deposit of different particles produced by a neutrino interaction,
we treat all the energy deposition at the same point source.
A new fitter for reconstructing neutrino interaction points and end points of the charged particle is currently under development.
\par
We select prompt events with visible energies ($E_{\text{prompt}}$) in the range of $50-1000$\,MeV, where the charge linearity of the PMTs and electronics has been confirmed by dye-laser calibration.
Furthermore, NCQE and CCQE interactions are dominant in this energy region.
We apply two spherical fiducial volume selection criteria with different radii:
A 450\,cm radius for $50 < E_{\text{prompt}} < 200 $\,MeV (low-E selection), and a 500\,cm radius for $200 < E_{\text{prompt}} < 1000$\,MeV (high-E selection).
Because fast neutron events are present as a background below 200\,MeV, we apply a tighter radius cut for the low-E selection.
Detailed information about the fast neutron background is described in Sec~\ref{sec:Simulation}.
We also apply OD cuts using the number of hit OD PMTs within a 200\,ns time window $N_{200\text{OD}}$ to cut cosmic muon backgrounds:
$N_{200\text{OD}}<5$ for periods I$-$III and $N_{200\text{OD}}<9$ for period IV.
Since we refurbished the OD system before the beginning of period IV, we adjust the threshold, so that veto efficiencies are equal.
The OD cuts reject atmospheric neutrino events where the final-state particles exit the ID.
All the events selected in this analysis are fully contained in the ID.
Overall, we find 425 events for the high-E selection and 114 events for the low-E selection.
The event rate in each period is stable within statistical errors.
\par
We select delayed events, {\it i.e.}, neutron capture gamma rays, using the delayed coincidence method.
We use the radius ($R_{\text{delayed}}$), the time difference from the prompt event ($\Delta T$), and the number of hit 17 inch PMTs within a 125\,ns time window ($NsumMax$).
We set $R_{\text{delayed}}<600$\,cm, which is well inside the LS region ($R<650$\,cm).
Immediately after a high-charge event, PMT afterpulses cause many noise events.
The high event rate leads to channel-level electronics deadtime effects, and many PMT waveforms are not recorded, making accurate energy reconstruction difficult.
Thus, we set $10 < \Delta T< 1000\,\mu$s and exclude events with a time delay less than 10\,$\mu$s.
We select delayed events using $NsumMax$ instead of the visible energy as it is less affected by these issues.
We set $NsumMax>275$ hits, a sufficiently low threshold to detect 2.2\,MeV gamma rays.
\par
\begin{figure}[htb]
\includegraphics[width=1.0\columnwidth]{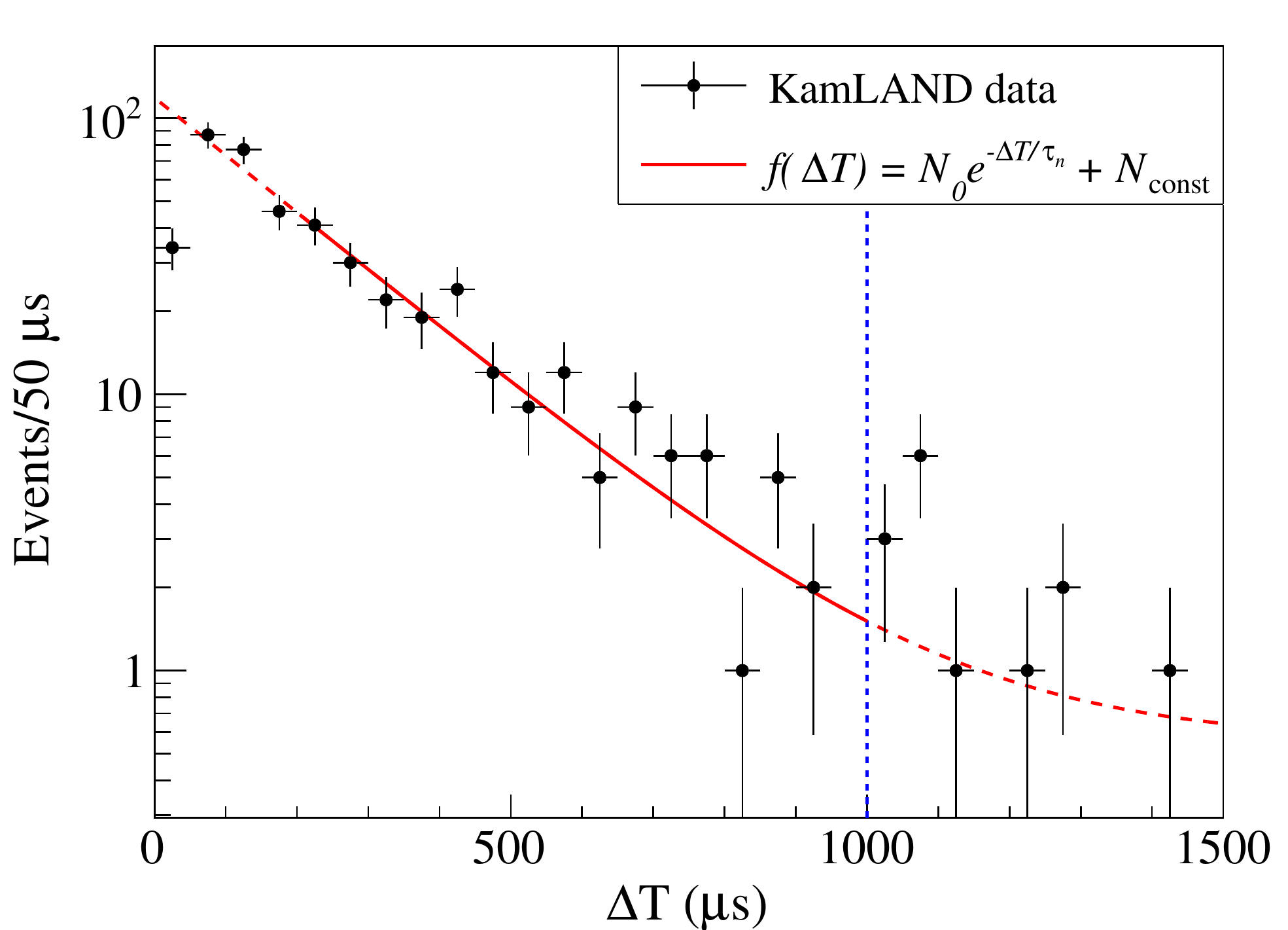}
\caption{Time difference between an atmospheric neutrino event (prompt) and a neutron capture event (delayed).
         All KamLAND atmospheric neutrino data sets are shown: both high-E and low-E selections during periods I-IV.
         The red line represents the fit result by Eq.~(\ref{eq:dT}) in the region $200 < \Delta T< 1000\,\mu$s.
         The blue dashed line represents the selection criteria corresponding to $\Delta T < 1000\,\mu$s.
         }
\label{fig:dT_candidate}
\end{figure}

Figure~\ref{fig:dT_candidate} shows the time difference between atmospheric neutrino events (prompt) and neutron capture events (delayed).
The detection inefficiency that occurs for $\sim50\,\mu$s immediately after atmospheric neutrino interactions can clearly be seen.
The $\Delta T$ distribution is fitted with a function,
\begin{align} \label{eq:dT}
 f(\Delta T) = N_0 e^{-\Delta T/\tau_n} + N_{\text{const}},
\end{align}
between $200 < \Delta T< 1000\,\mu$s, where $\tau_n= 207.5\,\mu$s.
The constant term $N_{\text{const}}$ corresponds to the background contamination in delayed events.
It is consistent with zero within a large uncertainty: $N_{\text{const}}=0.56\pm0.74$\,events/50\,$\mu$s.
The background event rate is also estimated using a long off-time window ($2 < \Delta T< 3002$\,ms).
The result is $(3.61\pm0.08) \times 10^{-2}$\,events/50\,$\mu$s.
This low event rate means we have negligible contamination in the delayed events, $(0.160\pm0.003)$\%.
The neutron tagging efficiency $\epsilon$ can be calculated from the actual number of observed neutrons ($N_{\text{obs}}$) and the integral of the fit result,
\begin{align} \label{eq:eff}
  \epsilon = \frac{N_{\text{obs}} - \int_{0\mu s}^{1000\mu s} N_{\text{const}} dt}
            {\int_{0\mu s}^{1000\mu s} N_0 e^{-t/\tau_n} dt},
\end{align}
estimating the inefficiency caused by the channel-level electronics deadtime effects.
The selection inefficiency caused by radius cut is taken into account in the detector simulation described in Sec.~\ref{sec:Simulation}.
The simulation also shows that the inefficiency associated with the gamma ray escaping the LS is insignificant.
We obtain $\epsilon=89.7^{+8.4}_{-7.3}$\% from Fig.~\ref{fig:dT_candidate}.
It is known that the neutron tagging efficiency in KamLAND has a prompt energy and a time dependence.
Since the leading causes of this inefficiency are afterpulses and the overshoots that occur $\sim 100$\,ns after a high-charge event,
these effects depend on the charge intensity of the prompt event.
In addition, PMT aging has gradually decreased the efficiency.
However, due to low statistics, the analysis performed here using neutrons associated with atmospheric neutrino events cannot evaluate the prompt energy and time dependence.
We therefore need an alternative way to estimate the efficiency more precisely.
A more precise analysis using cosmic muons is described in Sec.~\ref{sec:neutron_tag}.

\begin{figure}[htb]
\includegraphics[width=1.0\columnwidth]{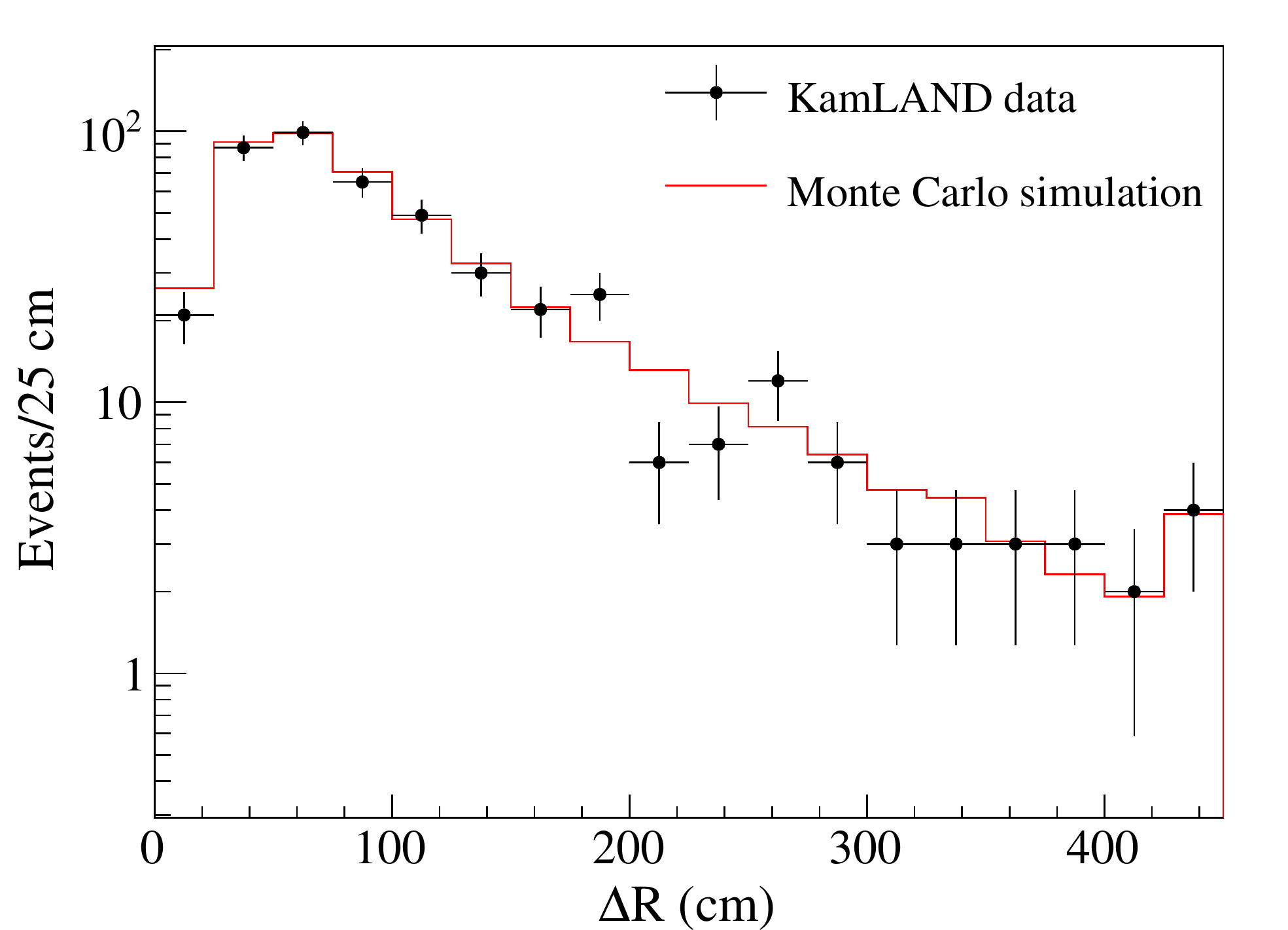}
\caption{Spatial separation between atmospheric neutrino events and neutron capture events.
         All KamLAND atmospheric neutrino data sets are shown, as in Fig.~\ref{fig:dT_candidate}.
         The red line represents the result of the Monte Carlo simulation before spectral fitting; the simulation assumes $M_A=1.2$\,GeV and $g_A^s=0$.
         The rightmost bin includes overflow.
         The simulation reproduces KamLAND data well.
         }
\label{fig:dR}
\end{figure}

Figure~\ref{fig:dR} shows the spatial difference between the prompt atmospheric neutrino interaction and delayed neutron capture events.
The spatial difference $\Delta R$ is the distance between the reconstructed positions of the center of energy deposition for the prompt and delayed events.
Since the neutrons emitted via the neutrino interaction have high energy, the $\Delta R$ distribution spreads widely.
The KamLAND data are compared with Monte Carlo simulation without any spectral fitting, with $M_A=1.2$\,GeV and $g_A^s=0$.
The Monte Carlo simulation and KamLAND data are in good agreement.
This consistency indicates that the Geant4 neutron transport model, used in the detector simulation, reproduces the data very well.
The simulation details are described in Sec.~\ref{sec:Simulation}.

After cuts, we find 356 delayed events in the high-E selection and 91 delayed events in the low-E selection, with negligible background contamination.
Note that the presence or absence of delayed events is irrelevant to the selection of prompt events.

\subsection{Neutron tagging efficiency} \label{sec:neutron_tag}
\begin{figure}[htb]
\centering
\includegraphics[width=1.0\columnwidth]{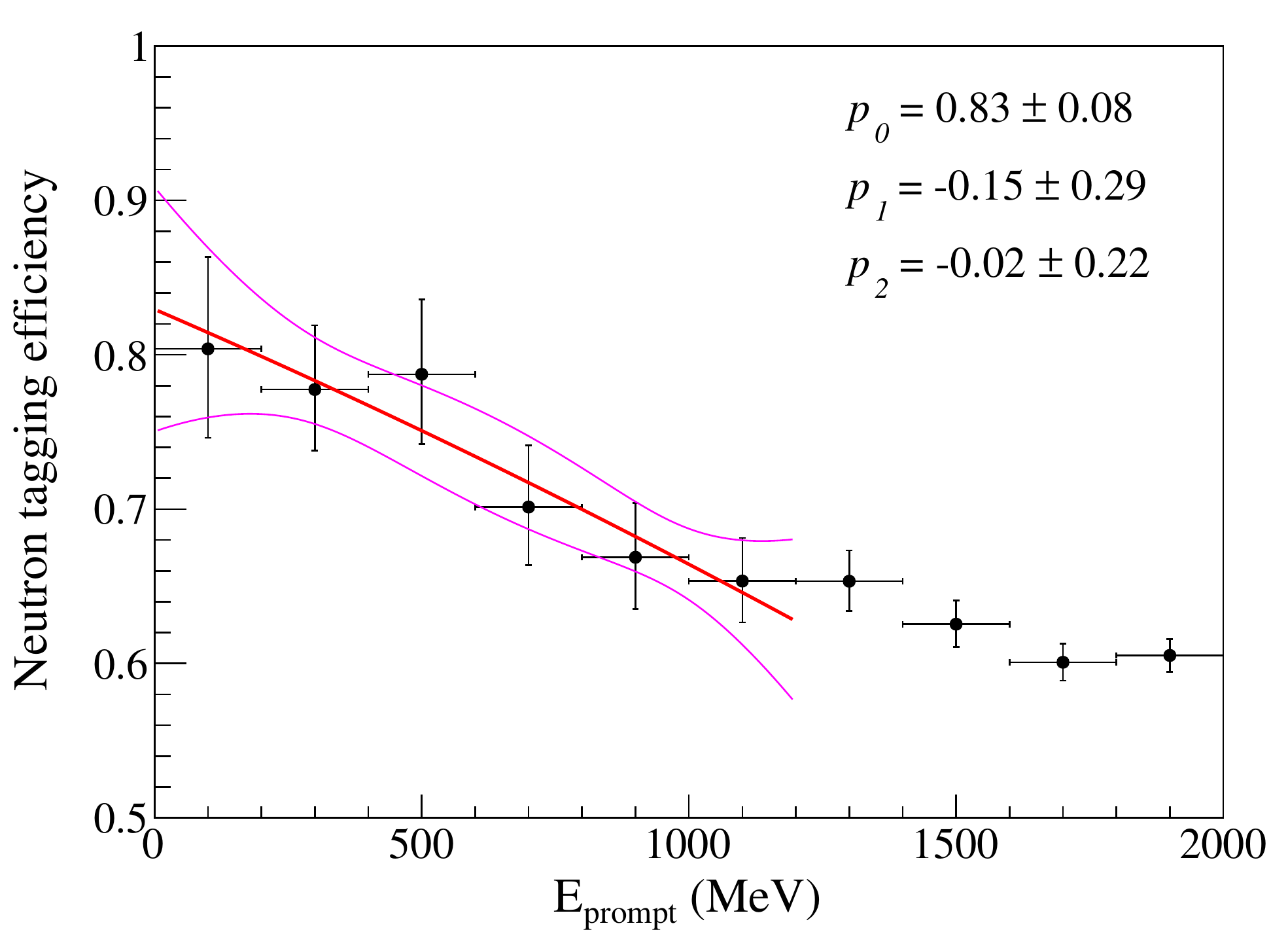}
\caption{Neutron tagging efficiency as a function of prompt visible energy during period IV.
         The red (magenta) lines represent the best fit ($1\sigma$ uncertainty) of the parametrization with the second-order polynomial of Eq.~(\ref{eq:eff_model}).
         Period IV has the lowest efficiency of the four periods.
         }
\label{fig:eff}
\end{figure}

As mentioned in Sec.~\ref{sec:selection}, the neutron tagging efficiency in KamLAND has prompt energy and time dependence.
For a more precise analysis, we parametrize the neutron tagging efficiency as a function of prompt energy for each period.
We use cosmic muons with high statistics as prompt events and apply the same selection criteria for the delayed events as in Sec.~\ref{sec:selection}.
The method of calculating the neutron tagging efficiency is the same.
For each prompt energy bin, the $\Delta T$ distribution is fitted with the function of Eq.~(\ref{eq:dT}).
The obtained $\Delta T$ distributions are similar to Fig.~\ref{fig:dT_candidate}, but differ in shape in the region $\Delta T < 150\,\mu$s, where the inefficiency occurs.
Using the fit results, we calculate the neutron tagging efficiency according to Eq.~(\ref{eq:eff}).
We confirm that the efficiency monotonically decreases over the experimental livetime of KamLAND and the prompt energy, within statistical uncertainty, as expected.
Figure~\ref{fig:eff} shows the efficiency obtained as a function of prompt energy for period IV.
The uncertainty is smaller than that obtained in Sec.~\ref{sec:selection} due to higher statistics.
The energy dependence is parametrized with a second-order polynomial for each period,
\begin{align} \label{eq:eff_model}
  \epsilon(E_{\text{prompt}}) = p_0 + p_1 E_{\text{prompt}} + p_2 E_{\text{prompt}}^2,
\end{align}
where $E_{\text{prompt}}$ has units of GeV.
The efficiency averaged over period I$-$IV is about 80\% at $E_{\text{prompt}}=1$\,GeV and 88\% at $E_{\text{prompt}}=0.1$\,GeV.
These values are consistent with the result obtained in Sec.~\ref{sec:selection}, $\epsilon=89.7^{+8.4}_{-7.3}$\%.
To take into account the prompt energy and time dependence of the efficiency, we use the values of $p_0,$ $p_1,$ $p_2$, and error matrices under this parametrization in the fits to energy spectra described in Sec.~\ref{sec:Analysis_Results}.

\section{Monte Carlo Simulation} \label{sec:Simulation}
Atmospheric neutrino events at KamLAND are estimated using Monte Carlo simulations.
We use the atmospheric neutrino flux calculations of HKKM 2014 above 100\,MeV~\cite{PhysRevD.92.023004} and Battistoni {\it et al.} below 100\,MeV~\cite{BATTISTONI2005526}.
While seasonal variations of the flux are negligible, less than 1\%, the effects of the solar cycle are not.
We discuss this effect in Sec.\ref{sec:solar_cycle}.
We calculate the neutrino oscillation effect propagating through the Earth using the Prob3++ package developed by members of Super-Kamiokande Collaboration~\cite{GitHubro82:online}.
The atmosphere is modeled as a vacuum.
The Earth is modeled as a sphere of radius 6371 km with a simplified version of the preliminary reference Earth model (PREM)~\cite{DZIEWONSKI1981297}.
This simplified version of PREM has four layers with a spherical density profile.
We use the three-flavor oscillation parameters assuming the normal hierarchy from~\cite{PhysRevD.98.030001}.
Note that the neutrino oscillation affects only CC interaction, namely the background of this analysis.
The uncertainty on the neutrino oscillation parameters gives no perceptible change in the sensitivity of this analysis.
\par
We use NuWro version 21.09 to simulate neutrino interactions.
For CCQE and NCQE interactions, the Llewellyn-Smith formalism with BBBA05 vector form factors is adopted.
Resonant pion-production (RES) processes are simulated with the Adler-Rarita-Schwinger formalism~\cite{ADLER1968189,PhysRevD.12.2644} with dipole form factors~\cite{PhysRevD.80.093001}.
The cross section discussed in this model has a 10\% normalization uncertainty.
For the 2p2h interaction, we choose the transverse enhancement model (TEM)~\cite{Bodek2011}, which is the only NC 2p2h model available in NuWro.
This model has a 20\% normalization uncertainty on the cross section.
The TEM model does not predict the fraction of $np$ pair targets in 2p2h interaction.
In electron scattering, experiments have confirmed $np$ pairs are dominant, with measured fractions of as $0.96^{+0.04}_{-0.22}$~\cite{doi:10.1126/science.1156675} and $0.92^{+0.08}_{-0.18}$~\cite{PhysRevLett.97.162504},
but the case of pure weak interactions is uncertain.
The theoretical calculation for the weak interaction predicts 67\% for the fraction of $np$ pairs~\cite{PhysRevD.88.113007}.
Based on the mean and deviation of these values, we assign $0.85_{-0.20}^{+0.15}$ in this study.
The cross section data used to model nucleon FSI is a custom fit model, which improves the agreement with the experimental data~\cite{PhysRevC.100.015505}.
The one used in pion FSI is based on~\cite{SALCEDO1988557}.
We use the local Fermi gas model, which is more accurate than the relativistic Fermi gas model.
\par
After the neutrino interaction and nuclear deexcitation simulations, the detector response is simulated using a Geant4-based Monte Carlo simulation called KLG4.
KLG4, which uses Geant4 version 9.6.p04, is a full optical detector simulation, including detailed descriptions of the KamLAND geometry and optical parameters.
We adopted a hadron physics package called ``QGSP\_BIC\_HP'', suitable for sub-GeV hadronic interaction and precise thermal neutron transportation.
The optical parameters, such as the light yield, quenching effect, and attenuation length, are tuned to reproduce the KamLAND data.
We primarily used radioactive source calibration data for tuning, including $^{60}$Co, $^{68}$Ge, and $^{137}$Cs sources.
The quenching effect is parametrized by Birk's formula~\cite{Birks_1951}.
The energy peaks of these sources agree within 3.5\%, and the vertex bias is less than 3\,cm.
We also estimate the energy scale uncertainty using spallation products of cosmic muons, $^{12}$B and $^{12}$N.
Using the energy spectra of these $\beta$ decays, which have endpoints at around 15\,MeV,
the uncertainty is estimated to be almost equal to that of the source calibration data.
Finally, we checked the charge scale uncertainty for the high-energy region using minimum-ionizing cosmic muons.
The charge peak of minimum ionization agrees within 8\%, and the value is used in the fit described in Sec.~\ref{sec:Analysis_Results}.
It is known that Birk's formula does not properly describe the quenching effects for heavier charged particles such as protons.
The proton quenching effect of KamLAND LS is precisely measured using a monochromatic neutron beam~\cite{YOSHIDA2010574}.
The quenching factor obtained by the experiment is parametrized by a formula proposed by Chou~\cite{PhysRev.87.904} that empirically extends Birk's formula.
KLG4 implements Chou's formula to describe the quenching effect for protons.
\par
Fast neutrons induced by cosmic muons in the surrounding rock and water are a dominant background below 200\,MeV.
The neutrons scatter on protons and carbon nuclei in the LS, mimicking prompt events.
Then, they are thermalized and captured on protons and carbon, creating delayed events.
Neutrons produced outside the detector are exponentially attenuated by the shielding layers of water, BO, and LS.
However, a contribution remains within the fiducial volume of this analysis.
We estimate the fast neutron background using the KLG4 and cosmic muon profile at the KamLAND site~\cite{PhysRevC.81.025807,Abe_2022}.
The uncertainty depends on the neutron production yield in rock, and the simulation takes considerable computation.
We conservatively assign a 100\% uncertainty to our estimate.

\subsection{Effect of the solar cycle}\label{sec:solar_cycle}
HKKM 2014 provides atmospheric neutrino flux data at the solar minimum and maximum.
The minimum and maximum are defined using the count rate of a specific neutron monitor (NM), the Climax NM~\cite{ClimaxNe33:online}.
This parameter is widely used to characterize the degree of solar activity.
There is a linear and inverse correlation between these parameters.
It is assumed that while the correlation gradient will depend on the location of various NMs, a linear correlation applies.
HKKM defines $4150$\,counts/hour/100 as the solar minimum, and $3500$\,counts/hour/100 as the solar maximum.
From the Climax NM data trend, we can adequately consider the solar cycle's effect on the atmospheric neutrino flux.
However, because the Climax NM was shut down in 2006, we need to calculate an equivalent Climax NM count, termed the NM parameter, using other NM data.
\par
\begin{figure*}[htb]
\centering
\includegraphics[width=0.75\textwidth]{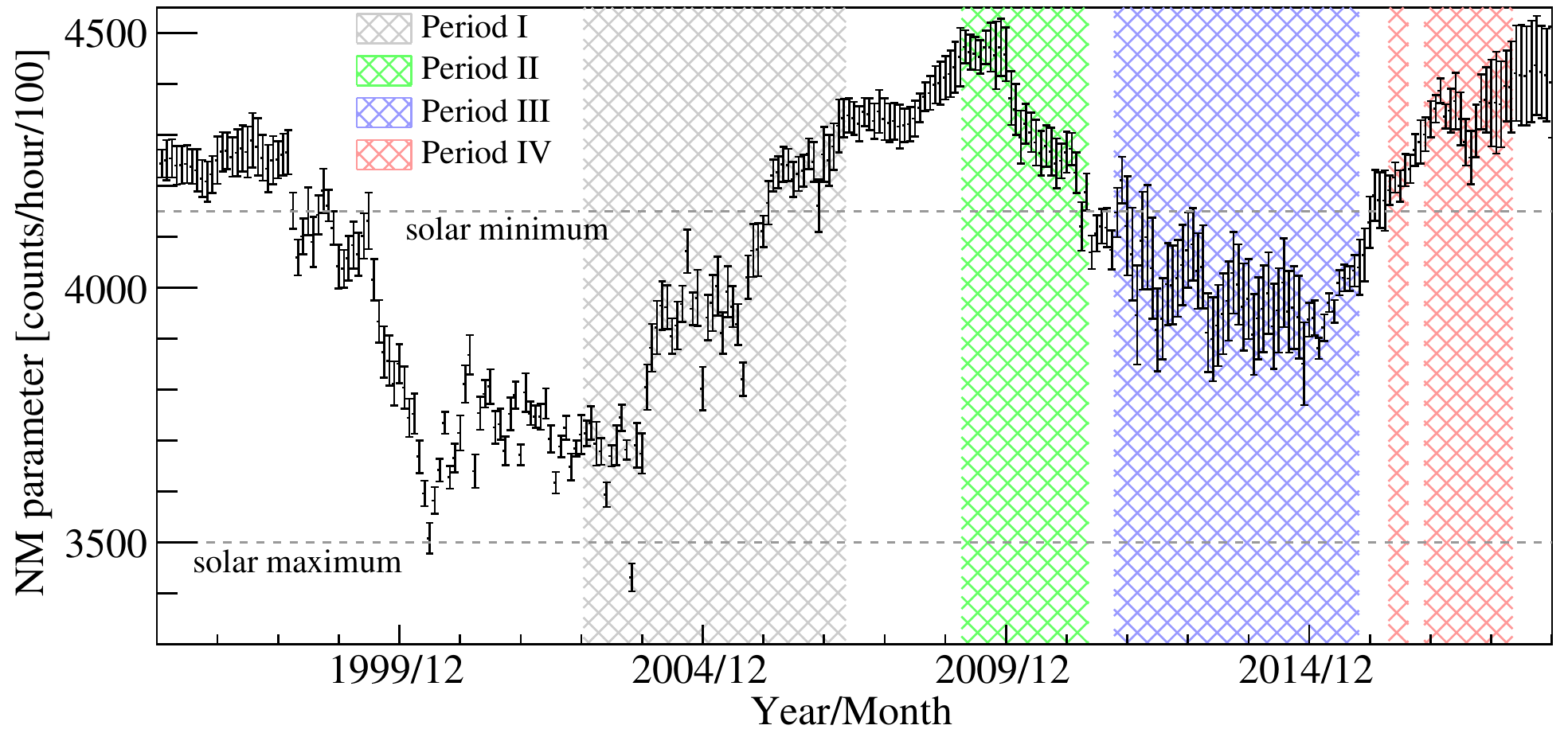}
\caption{Time variation of the NM parameter.
         The shaded regions denote the four analysis periods in this paper.
         The dashed lines represent the solar minimum and maximum as defined in the flux calculation of HKKM 2014~\cite{PhysRevD.92.023004}.
         The error bars are calculated from the standard deviation of the five converted count rates.
        }
\label{fig:solar_modulation}
\end{figure*}

We use five NM datasets in addition to the Climax NM, which cover the entire analysis period: the Moscow, Apatity, Thule, Newark, and Oulu neutron monitors~\cite{LinkstoC63:online,OuluCosm68:online}.
Their count rates have a linear correlation with the Climax NM data.
We fit the correlation between each dataset and the Climax NM with a first-order polynomial during the period for which both monitors were available.
We then convert the count rate of each monitor to the NM parameter, which is directly comparable to the Climax NM count rate.
Figure~\ref{fig:solar_modulation} shows the trend of the NM parameter.
Our data set indicates that solar cycle 24 had low solar activity.
This result is consistent with that obtained by the Super-Kamiokande Collaboration~\cite{PhysRevD.94.052001}.
We calculate the livetime-averaged NM parameter for each period as shown in Table~\ref{tab:NM_param}.
The relative normalization change due to the solar cycle is calculated to be about 3\%.
The uncertainty of these count rates is 110\,counts/hour/100 from the standard deviation of the five converted count rates.

\begin{table}[htb]
\centering
\caption{Livetime-averaged NM parameter.
         The HKKM defines 4150\,counts/hour/100 as the solar minimum, and 3500\,counts/hour/100 as the solar maximum.}
\label{tab:NM_param}
\begin{tabular}{lc} \hline \hline
Period & NM parameter (counts/hour/100) \\ \hline
Period I  & 3973 \\
Period II & 4327 \\
Period III & 3991 \\
Period IV  & 4318 \\ \hline \hline
\end{tabular}
\end{table}

\section{Analysis and Results} \label{sec:Analysis_Results}
\begin{figure}[htb]
\centering
\includegraphics[width=0.85\columnwidth]{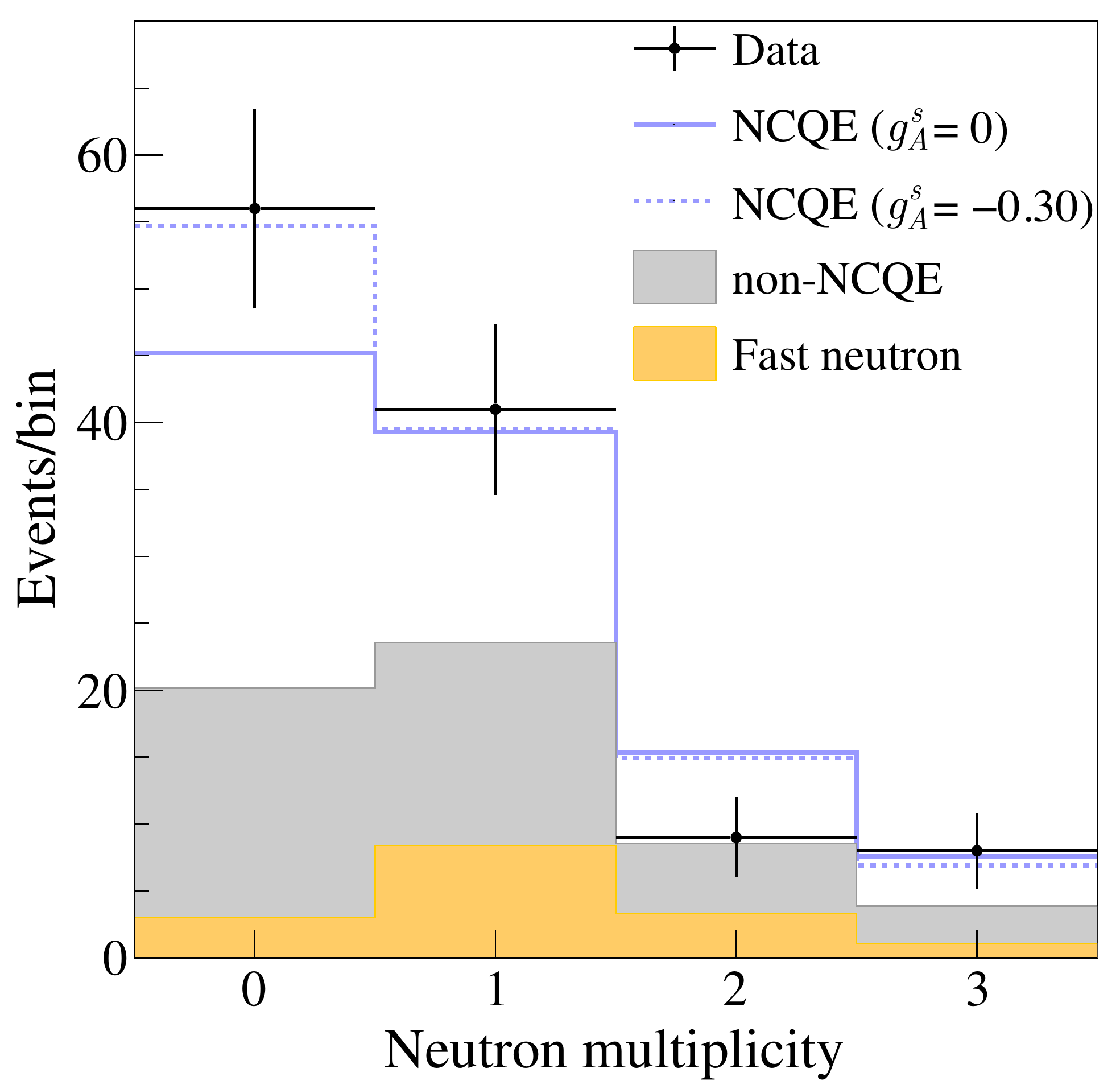}
\caption{Neutron multiplicity of atmospheric neutrino events in the low-E selection ($50\,\text{MeV} < E_{\text{prompt}} < 200$\,MeV).
         The orange-shaded region represents the expected fast neutron background, the gray-shaded region shows expected atmospheric neutrino events from interaction modes other than NCQE, and the blue solid (dashed) lines denote NCQE interactions with $g_A^s = 0~(-0.30)$.
         The rightmost bin includes overflow.
         The simulation data are shown prior to the spectral fit, assuming $M_A=1.2$\,GeV.
        }
\label{fig:nmulti_prefit}
\end{figure}

Before discussing the fits to event energy spectra, we briefly introduce how $g_A^s$ affects the KamLAND data.
The effect of $g_A^s$ appears as a change in the distribution of neutron multiplicities, while no apparent change is seen in the visible energy distribution.
Since the NCQE interaction is dominant below 200\,MeV, corresponding to the low-E selection in this analysis, the neutron multiplicity in that region is sensitive to $g_A^s$.
Figure~\ref{fig:nmulti_prefit} shows the neutron multiplicity distribution of atmospheric neutrino events in the low-E selection ($50\,\text{MeV} < E_{\text{prompt}} < 200$\,MeV).
Since negative $g_A^s$ increases the NCQE cross section with protons, the total cross section with KamLAND LS with its CH$_2$ composition also increases.
The NCQE interaction with free protons is not accompanied by neutron emission via FSI and nuclear deexcitation and typically leads to zero neutron multiplicity.
Thus, negative $g_A^s$ enhances the rate of NCQE events with zero neutron multiplicity.
Based on these considerations, we emphasize the importance of considering neutron multiplicity in the analysis.

We simultaneously extract $M_A$ and $g_A^s$ from a fit of visible energy spectra.
We used a binned $\chi^2$ method incorporating systematic uncertainties.
The $\chi^2$ is composed of a Poisson term $\chi^2_{\text{Poisson}}$ and a penalty term $\chi^2_{\text{penalty}}$:
\begin{align} \label{eq:chi2}
  \chi^2 = \chi^2_{\text{Poisson}} + \chi^2_{\text{penalty}}.
\end{align}
The Poisson term is defined using the number of observed events $n_{ijk}$ and the number of expected events $\nu_{ijk}$:
\begin{align} \label{eq:chi2_poisson}
  & \chi^2_{\text{Poisson}} =  \nonumber \\
  &
  \begin{cases}
    2\sum\limits_i\sum\limits_j\sum\limits_k\left[\nu_{ijk} - n_{ijk}\right] & (n_{ijk}=0) \\
    2\sum\limits_i\sum\limits_j\sum\limits_k\left[\nu_{ijk} - n_{ijk} +n_{ijk}\log(n_{ijk}/\nu_{ijk})\right] & (n_{ijk}>0),
  \end{cases}
\end{align}
where the indices $i,j,$ and $k$ represent the $i$th period, $j$th visible energy, and $k$th neutron multiplicity bins.
We have four data collection period bins corresponding to periods I-IV.
We also have thirteen visible energy bins, eight for the high-E selection and five for the low-E selection.
We divide the data into four neutron multiplicity bins, neutron multiplicity 0, 1, 2, and 3 or more.
The analysis can consider neutron multiplicity by including the neutron multiplicity bins in the Poisson term.
The penalty term is defined as
\begin{align} \label{eq:chi2_penalty}
  \chi^2_{\text{penalty}} = &\sum\limits_l \left( \frac{E_l-O_l}{\sigma_l}\right)^2 \nonumber \\
                            & + \sum\limits_n \sum\limits_m (E_n-O_n) M_{nm}^{-1} (E_m-O_m),
\end{align}
where $l$ represents a systematic uncertainty parameter other than the neutron tagging efficiency, $E_l$ is the expected value, $O_l$ is the observed value in the fit,
and $\sigma_l$ is expected uncertainty of the parameter $l$.
The indicides~$n$ and $m$ denote parameters of the neutron tagging efficiency, and $M_{nm}^{-1}$ represents the inverse of error matrix described in Sec.~\ref{sec:neutron_tag}.
The systematic uncertainties considered in this analysis are summarized in Tables~\ref{tab:syst_common}, \ref{tab:syst_deex}, and \ref{tab:syst_independent}.

\subsection{Results and discussion}
\begin{figure*}[htbp]
\centering
\includegraphics[width=0.90\textwidth]{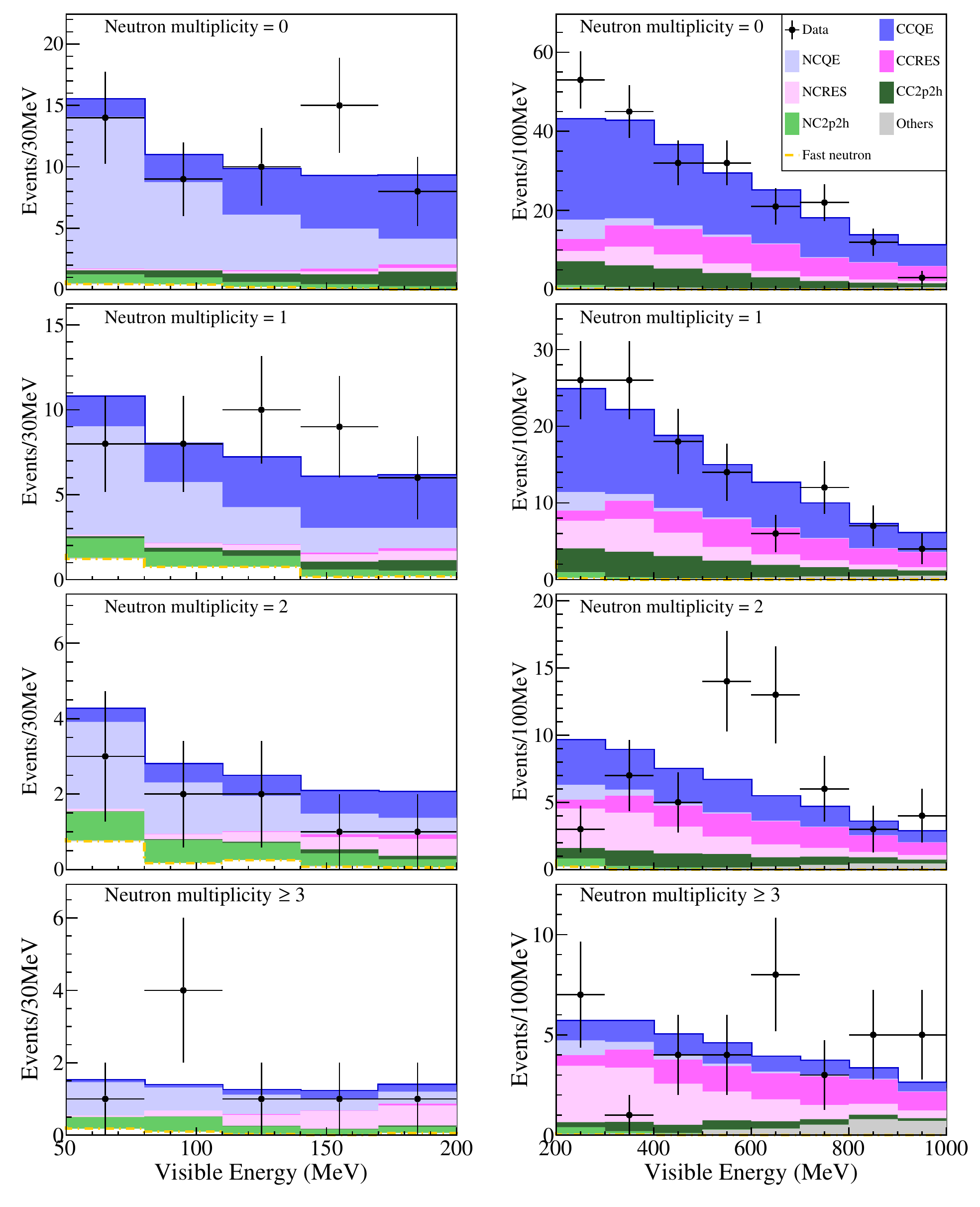}
\caption{Best-fit visible energy spectra with neutron multiplicity 0, 1, 2, and $\geq 3$.
         The left figures show the low-E selection, between 50 and 200\,MeV.
         The right figures show the high-E selection, between 200\,MeV and 1000\,MeV.
         The ``others'' category in gray refers to deep-inelastic and coherent scattering.
         The NCQE interaction is dominant below 200\,MeV.
         }
\label{fig:bestfit}
\end{figure*}

\begin{figure}[htb] \centering
\includegraphics[width=1.0\columnwidth]{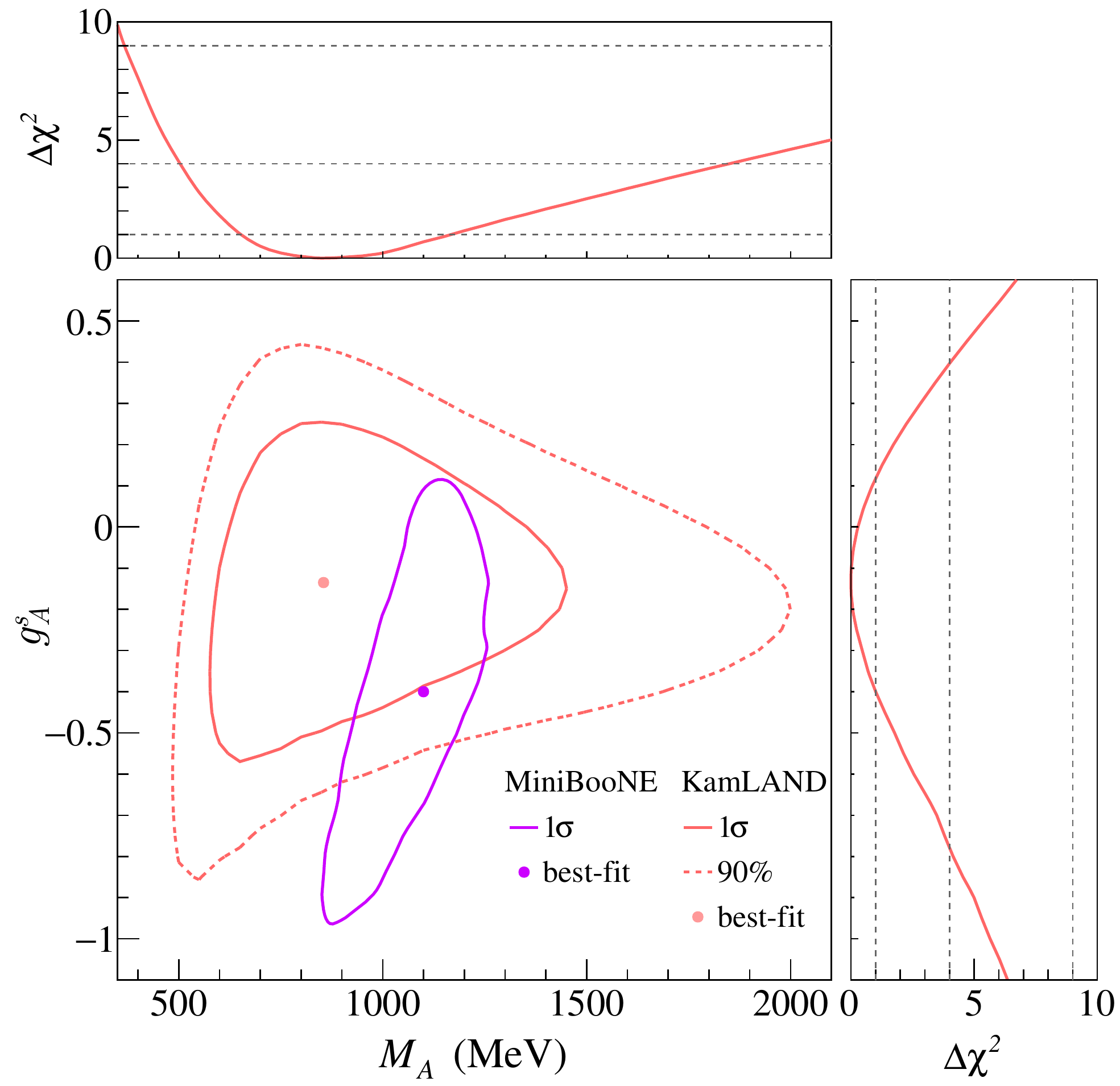}
\caption{Two-dimensional allowed regions for $M_A$ and $g_A^s$.
         The red contour and dot are the result of this work.
         The side panels show the one-dimensional $\Delta\chi^2$-profiles projected onto $M_A$ and $g_A^s$.
         The violet contour and dot display the~$1\sigma$~CL and best-fit value from Golan {\it et al.} using MiniBooNE data~\cite{PhysRevC.88.024612}.
         The parameter $M_A$ is treated as a free parameter in both results.
         }
\label{fig:dchi2}
\end{figure}

Figure~\ref{fig:bestfit} shows the best-fit visible energy spectra.
The KamLAND data, which measure the neutron multiplicity with almost 80\% efficiency, are well described by the simulations over a wide energy range, $50-1000$\,MeV.
The NCQE interaction dominates in the low-energy region, roughly below 200\,MeV.
The neutron multiplicity in this energy region determines the value of~$g_A^s$.
Figure~\ref{fig:dchi2} shows the two-dimensional allowed regions for $M_A$ and $g_A^s$.
We obtain $M_A = 0.86^{+0.31}_{-0.20}$\,GeV and $g_A^s = -0.14^{+0.25}_{-0.26}$.
Our result is consistent with the result by Golan {\it et al.} using MiniBooNE data~\cite{PhysRevC.88.024612}.
The plot shows little dependence on $M_A$, as expected.
This feature, realized by measuring neutron multiplicity, is important in the present experimental situation where measured values of $M_A$ vary from experiment to experiment.

\begin{figure}[htb] \centering
\includegraphics[width=1.0\columnwidth]{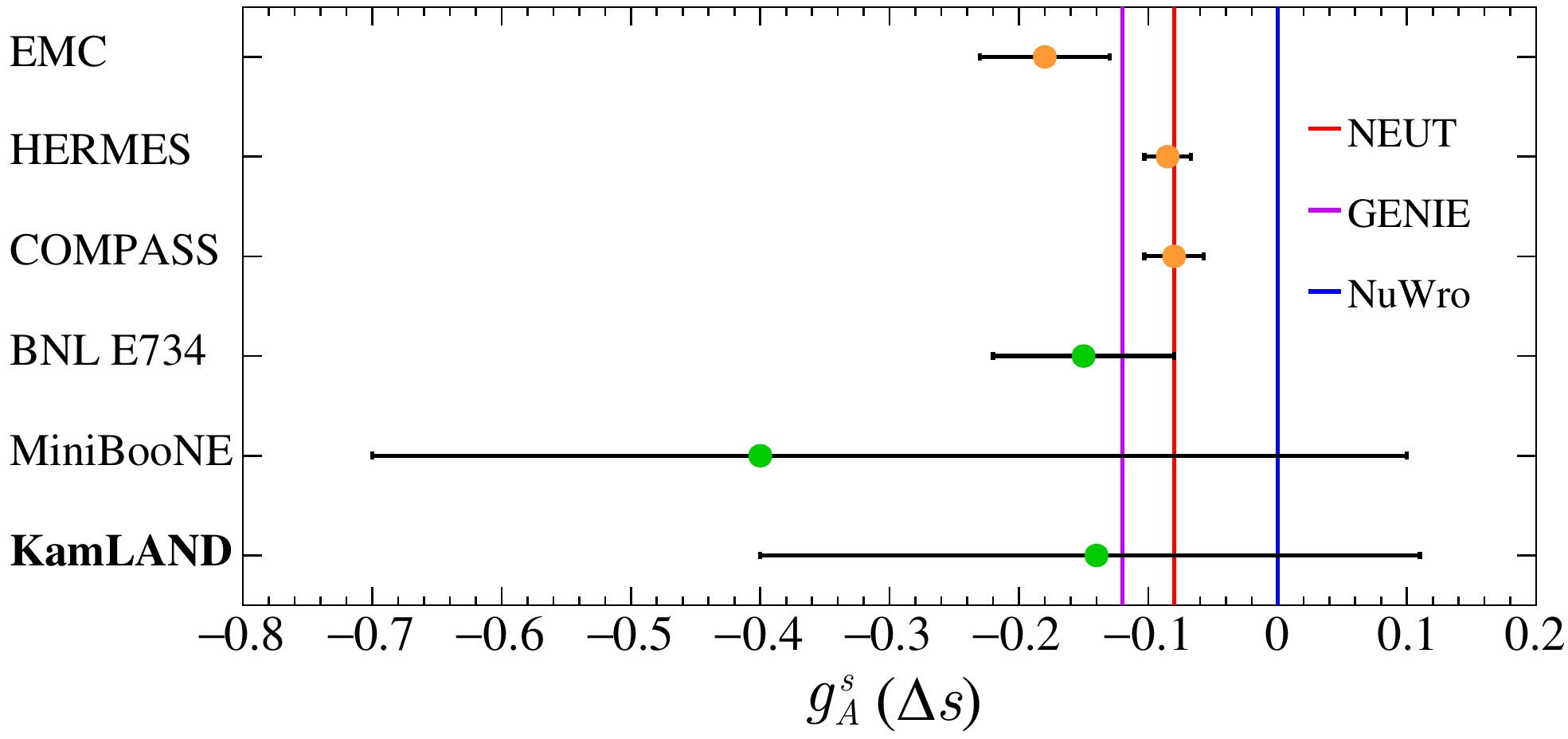}
\caption{Summary of $g_A^s$ ($\Delta s$) measurements.
         In addition to the result of this work, results from EMC~\cite{ASHMAN19891,ALBERICO2002227}, HERMES~\cite{PhysRevD.75.012007}, COMPASS~\cite{ALEXAKHIN20078}, BNL E734~\cite{PhysRevC.48.761}, and Golan {\it at al.} using MiniBooNE data~\cite{PhysRevC.88.024612} are shown.
         Results with orange symbols are polarized-lepton deep-inelastic scattering experiments, and those with green symbols are neutrino NCQE scattering experiments.
         The red, violet, and blue vertical lines represent the default values adopted in several neutrino Monte Carlo generators.
        }
\label{fig:gAs_summary}
\end{figure}

A summary of $g_A^s (\Delta s)$ measurements and the values adopted in neutrino Monte Carlo event generators is shown in Fig.~\ref{fig:gAs_summary}.
All the experimental results have consistent values and prefer a negative value of $g_A^s$.
A negative $g_A^s$ is reasonably explained by the current experimental measurements of hadronic matrix elements~\cite{PhysRevD.67.037503}.
We should note two points in the interpretation of Fig.~\ref{fig:gAs_summary}.
The first is that the impact of $SU(3)_f$ flavor symmetry breaking on polarized-lepton deep-inelastic scattering experiments is not included.
The second point concerns the treatment of $M_A$ and the 2p2h contribution.
As described in Sec.~\ref{sec:Introduction}, it is difficult to determine a reasonable constraint on $M_A$ in the current experimental situation.
The MiniBooNE and KamLAND results were obtained without $M_A$ constraints and included consideration of the 2p2h interaction.
In contrast, the BNL~E734 result did not consider any 2p2h interaction, and $M_A$ was strongly constrained.
The BNL E734 result could therefore be affected by the contribution of the 2p2h interaction and a larger $M_A$ uncertainty.
Our result gives the most stringent limit on $g_A^s$ among NCQE measurements without $M_A$ constraints.
The experimental NCQE data prefer smaller values than the results of polarized-lepton deep-inelastic scattering experiments and those adopted in neutrino Monte Carlo generators.
However, they still have large uncertainties and are not yet accurate enough to claim adequate theoretical inputs.
Further improvements in both experimental accuracy and theoretical modeling will be necessary.

In our analysis, the systematic and statistical uncertainties on $g_A^s$ are almost comparable: $g_A^s = -0.14 \pm 0.17 (\text{stat})^{+0.18}_{-0.20}$(syst).
The dominant systematic uncertainties come from FSI and the 2p2h interaction, followed by the KamLAND neutron tagging efficiency.
In order to improve the sensitivity to $g_A^s$, we need to optimize the FSI model.
Electron scattering data can validate it, but it is not easy to model the dynamics of strong interactions in nuclei.
Recently, S.~Dytman {\it et al.} reported detailed comparisons of the FSI models implemented in NuWro, NEUT, and GENIE in~\cite{PhysRevD.104.053006}.
They show significant variations between the generators, and further discussion is necessary.
We also need to measure the 2p2h interaction directly and check the model's validity.
Direct measurements of 2p2h interactions using detectors with good track resolution are planned and ongoing~\cite{PhysRevD.102.072006}.
A combined analysis with the data from those experiments will be able to constrain $g_A^s$ further.
Nieves {\it et al.}~\cite{PhysRevC.83.045501} and Martini {\it et al.}~\cite{PhysRevC.80.065501} have been developing microscopic models to describe the 2p2h interaction.
However, since they mainly focus on the CC, only the TEM is currently available for NC 2p2h in the generators.
This situation has forced us to rely on the TEM in this analysis.
The TEM is a theoretically effective model and relies on electron scattering experiments.
The model uncertainties calculated in deriving the TEM are accurately considered in this analysis.
Since the 2p2h interaction is due to nuclear effects, it is natural to consider an analogy with electron scattering experiments.
Nevertheless, verification by direct measurement is required.
We also expect that NC 2p2h models other than the TEM will be developed and implemented into the generators to allow verification of various models.

\section{Conclusion and Prospects} \label{sec:Conclusion}
We report a new measurement of the strange axial coupling constant $g_A^s$ using neutron multiplicity associated with the NCQE interaction of atmospheric neutrino at KamLAND.
A simulation method for nuclear deexcitation, which is required to predict the neutron multiplicity accurately, is established.
We use KamLAND atmospheric neutrino data from January 2003 to May 2018, corresponding to 10.74\,years of total live time.
By fitting the visible energy spectrum for each neutron multiplicity, we obtain $g_A^s = -0.14^{+0.25}_{-0.26}$, which is the most stringent limit obtained using NCQE interactions without $M_A$ constraints.
The experimental data on NCQE interactions, including this result, favor slightly smaller values than those used in the neutrino Monte Carlo generators.
However, further improvements in accuracy are necessary to claim an appropriate value.
\par
The main future tasks for enhancing the accuracy are detailed investigations of the FSI models and a direct measurement and model validation of 2p2h interaction.
We need careful investigations on the FSI models to understand the large differences among the generators and to optimize the models to give good consistency with experimental data.
Since only the TEM is currently available for NC 2p2h in the generators, this analysis has been forced to rely on the TEM, which relies on electron scattering experiments.
Validation of the TEM by directly measuring the 2p2h is a high-priority future task.
Although direct measurement of the 2p2h interaction is quite challenging at KamLAND,
a combined analysis with other experiments, which aim at the direct measurements of the 2p2h, will be effective.
We also expect that NC 2p2h models other than the TEM will be implemented into the generators.
\par
In recent neutrino physics, the importance of accurately determining $g_A^s$ and comprehensively predicting neutron multiplicity, including nuclear deexcitation, has increased dramatically.
Detectors capable of measuring neutron multiplicity have been rare, but recent and next-generation detectors, such as Super-Kamiokande Gadolinium~\cite{ABE2022166248}, Hyper-Kamiokande~\cite{https://doi.org/10.48550/arxiv.1805.04163}, and JUNO~\cite{An_2016}, will make it possible.
These experiments plan to use neutron tagging information to significantly reduce the main background, atmospheric neutrino events, in searches for supernova relic neutrinos and proton decay.
The dominant systematic uncertainty in these analyses comes from neutrino-nucleus interactions, especially the nuclear effects related to neutron emission.
Therefore, the prediction accuracy of neutron emission in neutrino interactions affects the accuracy of these observations.
Since the NCQE interaction of atmospheric neutrino is the main background to the supernova relic neutrino search, the determination of $g_A^s$ will be essential.
Next-generation detectors will significantly improve the measurement statistics, so reducing these systematic uncertainties is essential.
\par
This analysis is the first to measure neutron multiplicity with a detection efficiency of $\sim$80\%.
It is also the first to compare measured neutron multiplicity with simulations that consider nuclear deexcitation.
This analysis will add significant knowledge to the many recent and next-generation experiments mentioned above.
All of them must consider nuclear deexcitation processes when conducting these studies.
We expect to integrate the nuclear deexcitation simulation developed here into neutrino event generators for use in other experiments.

\begin{acknowledgments}
The \mbox{KamLAND} experiment is supported by JSPS KAKENHI Grant No. 19H05803, the Dutch Research Council (NWO), by the U.S. Department of Energy (DOE) Grant No.\,DE-AC02-05CH11231, by the National Science Foundation (NSF) Grant No.\,2021964, as well by as other DOE and NSF grants to individual institutions.
The Kamioka Mining and Smelting Company has provided services for our activities in the mine.
We acknowledge the support of NII for Science Information Network (SINET).
This work was supported by JSPS KAKENHI Grant No. 21J10185 and by the Graduate Program on Physics for the Universe (GP-PU), Tohoku University.
We also thank Y.~Hayato and K.S.~McFarland for advising us on our neutrino simulation.
\end{acknowledgments}

\appendix*
\section{SYSTEMATIC UNCERTAINTIES}\label{sec:systematics}
Tables~\ref{tab:syst_common}, \ref{tab:syst_deex} and \ref{tab:syst_independent} show the systematic uncertainties in this analysis and their best-fit values.

\begin{table*}[hp]
\centering
\caption{Systematic uncertainties related to the flux, cross section, final-state interactions and secondary interactions~(SI).
         These are common to all data collection periods.}
\label{tab:syst_common}
\begin{tabular}{llcc} \hline \hline
  \multicolumn{2}{l}{Parameter} & Expected & Best-fit \\ \hline
  Flux normalization & $E_\nu<0.1$\,GeV   & $1.00\pm0.35$ & $ 0.98\pm0.34 $ \\
                     & $0.1<E_\nu<1$\,GeV & $1.00\pm0.35$ & $ 1.51\pm0.11 $ \\
                     & $E_\nu>1$\,GeV     & $1.00\pm0.15$ & $ 0.98\pm0.11$ \\
  $\bar{\nu}_e/\nu_e$     &                   & $0.00\pm0.05$ & $0.00\pm0.05$ \\
  $\bar{\nu}_\mu/\nu_\mu$ &                   & $0.00\pm0.05$ & $0.00\pm0.05$ \\
  $(\nu_\mu + \bar{\nu}_\mu)/(\nu_e + \bar{\nu}_e)$ & & $0.00\pm0.02$ & $0.00\pm0.02 $ \\
  Cross section normalization & CCRES   & $1.00\pm0.10$ & $1.01\pm0.09 $ \\
                              & NCRES   & $1.00\pm0.10$ & $0.99\pm0.09 $ \\
                              & CC 2p2h & $1.00\pm0.20$ & $1.09\pm0.20$ \\ 
                              & NC 2p2h & $1.00\pm0.20$ & $0.98\pm0.19$ \\ 
  Fraction of $np$ pair target in 2p2h & & $0.85^{+0.15}_{-0.20}$ & $ 0.81\pm0.19$ \\
  Final-state interactions \footnote{Scale factors corresponding to the FSI probability} & nucleon & $1.00\pm0.28$ & $ 0.91\pm0.18$ \\
      & pion elastic    & $1.00\pm0.50$ & $1.09\pm0.40 $\\
      & pion absorption & $1.00\pm0.50$ & $1.08\pm0.44$ \\
  Secondary interaction  \footnote{Scale factors corresponding to the SI probability} & nucleon & $1.00\pm0.07$ & $ 1.00\pm0.07 $ \\
      & pion    & $1.00\pm0.14$ & $ 1.08\pm0.11 $ \\
  \hline \hline
\end{tabular}
\end{table*}

\begin{table*}[hp]
\centering
\caption{Systematic uncertainties related to the branching ratios of nuclear deexcitation from the $s_{1/2}$-hole state.
         These are common to all data collection periods.}
\label{tab:syst_deex}
\begin{tabular}{llcc} \hline \hline
  \multicolumn{2}{l}{Parameter} & Expected & Best-fit \\ \hline
  Single-step deexcitation of $^{11}$B$^*$ (\%) & Neutron  & $18.7_{-3.0}^{+6.1}$ & $18.5\pm2.3$ \\ 
                                                 & Proton   & $5.7_{-1.1}^{+2.9}$ & $5.7\pm1.3$ \\
                                                 & $\alpha$ & $3.3_{-0.5}^{+4.4}$ & $3.4\pm1.2$ \\
                                                 & Deuteron & $4.1_{-0.6}^{+4.1}$ & $4.1\pm0.8$ \\
                                                 & Triton   & $1.5_{-0.2}^{+14.9}$ & $1.6\pm1.2$ \\
  Multistep deexcitation of $^{11}$B$^*$ (\%)  & Neutron   & $25.3_{-7.6}^{+9.0}$    & $25.0\pm4.3$ \\
                                                 & Proton    & $1.4_{-0.2}^{+8.0}$     & $1.7\pm1.1$ \\
                                                 & $\alpha$  & $0.08_{-0.01}^{+12.04}$ & $0.08\pm0.08$ \\
                                                 & Deuteron  & $1.6_{-0.4}^{+3.3}$    & $1.6\pm0.5$ \\
                                                 & Triton   & $0.6_{-0.3}^{+1.7}$ & Fixed \\
  Single-step deexcitation of $^{11}$C$^*$ (\%) & Neutron  & $4.2_{-0.6}^{+1.7}$ & $4.2\pm0.9$ \\
                                                 & Proton   & $31.0\pm10.1$       & $30.9\pm5.0$ \\
                                                 & $\alpha$ & $8.9_{-1.3}^{+1.4}$ & $9.0\pm1.5$ \\
  Multistep deexcitation of $^{11}$C$^*$ (\%) & Neutron  & $1.5_{-0.2}^{+6.4}$  & $1.5\pm0.3$ \\
                                                & Proton   & $42.3\pm6.7$         & $42.3\pm6.3$ \\
                                                & $\alpha$ & $1.7_{-0.3}^{+10.0}$ & $1.7\pm0.3$ \\
  Neutron emission following multinucleon disappearance \footnote{Scale factors corresponding to the probabilities of neutron emission.}
  & Two nucleon disappearance & $1.0_{-1.0}^{+1.2}$ & $-0.4\pm0.7$  \\
  & Three or more nucleon disappearance & $1.0_{-1.0}^{+1.8}$ & $0.8\pm1.3$ \\
  \hline \hline
\end{tabular}
\end{table*}

\begin{table*}[htbp]
\caption{Systematic uncertainties related to that are independent in period I, period II, period III, and period IV.}
\label{tab:syst_independent}
\begin{minipage}[b]{1.00\hsize} \centering
\begin{tabular}{llcccc} \hline \hline
  \multicolumn{2}{l}{Parameter} & \multicolumn{2}{c}{Period I} & \multicolumn{2}{c}{Period II} \\
                               && Expected & Best-fit & Expected & Best-fit \\ \hline
  Solar cycle \footnote{Error factor corresponding to 110\,counts/hour/100 in NM parameter.}
      & & $0.00\pm1.00$ & $-0.01\pm0.96$ & $0.00\pm1.00$ & $0.03\pm0.99$\\
  Energy scale  & & $1.00\pm0.08$ & $1.04\pm0.07$ & $1.00\pm0.08$ & $0.92\pm0.07$ \\
  Fiducial volume  & Prompt  & $1.00\pm0.10$ & $0.99\pm0.07$ & $1.00\pm0.10$ & $1.02\pm0.08$ \\
                   & Delayed \footnote{Error factor corresponding to 0.7/0.4\% changes on the number of tagged neutrons in high-E/low-E selection.}
                             & $0.00\pm1.00$ & $0.02\pm1.01$ & $0.00\pm1.00$ & $-0.02\pm0.99$  \\
  Fast neutron normalization & & $1.00\pm1.00$ & $0.09\pm0.92$ & $1.00\pm1.00$ & $0.31\pm0.71$ \\
  Neutron tagging efficiency 
                             & $p_0$ & $0.90\pm0.06$  & $0.89\pm0.04$  & $0.91\pm0.06$  & $0.91\pm0.01$ \\
                             & $p_1$ & $0.13\pm0.22$  & $0.18\pm0.17$ & $-0.28\pm0.19$ & $-0.28\pm0.02$ \\
                             & $p_2$ & $-0.16\pm0.17$ & $-0.19\pm0.13$ & $0.16\pm0.14$  & $0.16\pm0.02$ \\
  \hline \hline
\end{tabular}
\end{minipage}
\begin{minipage}[b]{1.00\hsize} \centering
\vspace{10pt}
\begin{tabular}{llcccc} \hline \hline
  \multicolumn{2}{l}{Parameter} & \multicolumn{2}{c}{Period III} & \multicolumn{2}{c}{Period IV} \\
                               && Expected & Best-fit & Expected & Best-fit \\ \hline
  Solar cycle 
      & & $0.00\pm1.00$ & $-0.05\pm0.98$ & $0.00\pm1.00$ & $0.07\pm1.01$ \\
  Energy scale  & & $1.00\pm0.08$ & $0.95\pm0.06$ & $1.00\pm0.08$ & $1.00\pm0.07$ \\
  Fiducial volume  & Prompt  & $1.00\pm0.10$ & $0.96\pm0.07$ & $1.00\pm0.10$ & $1.08\pm0.08$ \\
                   & Delayed 
                             & $0.00\pm1.00$ & $0.13\pm0.99$ & $0.00\pm1.00$ & $-0.11\pm0.97$ \\
  Fast neutron normalization & & $1.00\pm1.00$ & $0.41\pm0.65$ & $1.00\pm1.00$ & $0.99\pm0.80$ \\
  Neutron tagging efficiency 
                             & $p_0$ &$0.86\pm0.04$  & $0.86\pm0.02$ & $0.83\pm0.08$  & $0.81\pm0.04$  \\
                             & $p_1$ & $-0.26\pm0.13$ & $-0.25\pm0.03$ & $-0.15\pm0.29$ & $-0.12\pm0.09$  \\
                             & $p_2$ & $0.11\pm0.09$  & $0.10\pm0.07$ & $-0.02\pm0.22$ & $-0.02\pm0.08$  \\
  \hline \hline
\end{tabular}
\end{minipage}
\end{table*}

\bibliography{AtmosphericNeutrino}

\end{document}